\documentclass[journal,10pt,twocolumn]{IEEEtran}
\newif\ifdouble
\doubletrue
\pdfoutput=1
\usepackage[utf8x]{inputenc}
\usepackage[]{standalone}
\usepackage{textcomp}
\usepackage{tikz,pgfplots}
\usepackage{graphicx}
\usepackage[caption=false]{subfig}
\usepackage{amsmath}
\usepackage{pgfplotstable}
\usepackage{booktabs}
\usepackage{multirow}
\usepackage{amsthm}
\usepackage{amssymb}
\usepackage{amsfonts}
\usepackage{wasysym}
\usepackage{algorithm}
\usepackage{algorithmic}
\usepackage{hyperref}
\usepackage{rotating}
\usepackage{url}
\usepackage{xcolor}
\usepackage{array}
\usepackage[numbers,sort&compress]{natbib}

\catcode`~=11

\catcode`~=13

\newcolumntype{L}[1]{>{\raggedright\let\newline\\\arraybackslash\hspace{0pt}}m{#1}}
\newcolumntype{C}[1]{>{\centering\let\newline\\\arraybackslash\hspace{0pt}}m{#1}}
\newcolumntype{R}[1]{>{\raggedleft\let\newline\\\arraybackslash\hspace{0pt}}m{#1}}
\usetikzlibrary{arrows,automata,calc,shapes,shapes.symbols,positioning,fit,shadows}

\pgfplotsset{
    compat=1.12,
    every legend to name picture/.style={scale=0.7},
    every legend to name picture/.style={font=\scriptsize},
}

\definecolor{mycolor1}{rgb}{0.85,0.85,1.0}
\definecolor{mycolor2}{rgb}{1.0,0.85,0.85}
\definecolor{mycolor3}{rgb}{0.0,0.75,0.75}%
\definecolor{mycolorg}{rgb}{0.5,1.00,0.5}
\definecolor{mycolorr}{rgb}{1.0,0.5,0.5}
\definecolor{mycolorm}{rgb}{1.0,0.0,1.0}%
\definecolor{green}{rgb}{0.0,0.5,0.0}
\definecolor{lightgreen}{rgb}{0.0,1.0,0.0}

\usepackage[disable,colorinlistoftodos,prependcaption,textsize=scriptsize]{todonotes}

\title{FiFTy: Large-scale File Fragment Type Identification using Convolutional Neural Networks}

\author{
  Govind Mittal, Paweł Korus,~\IEEEmembership{Member,~IEEE} and Nasir Memon,~\IEEEmembership{Fellow,~IEEE}
  \thanks{
  \textcopyright 2020 IEEE.  Personal use of this material is permitted.  Permission from IEEE must be obtained for all other uses, in any current or future media, including reprinting/republishing this material for advertising or promotional purposes, creating new collective works, for resale or redistribution to servers or lists, or reuse of any copyrighted component of this work in other works.}

  \thanks{G. Mittal and N. Memon are with the Tandon School of Engineering, New York University, USA. Part of this work was performed was supported by NYU Abu Dhabi, UAE. (e-mail: \{mittal,memon\}@nyu.edu).}
  \thanks{P. Korus is with the Tandon School of Engineering, New York University, USA, and also with the Department of Telecommunications, AGH University of Science and Technology, Poland (e-mail: pkorus@nyu.edu).}
  \thanks{Supplementary materials, inlcuding source code, all data sets, and additional results are freely available at \url{https://github.com/mittalgovind/fifty}}
  \vspace*{-0.5cm}
}

\begin{document}

\markboth{IEEE TRANSACTIONS ON INFORMATION FORENSICS AND SECURITY}{}

\maketitle

\begin{abstract}
We present \emph{FiFTy}, a modern file-type identification tool for memory forensics and data carving. In contrast to previous approaches based on hand-crafted features, we design a compact neural network architecture, which uses a trainable embedding space. Our approach dispenses with the explicit feature extraction which has been a bottleneck in legacy systems. We evaluate the proposed method on a novel dataset with 75 file-types -- the most diverse and balanced dataset reported to date. \emph{FiFTy} consistently outperforms all baselines in terms of speed, accuracy and individual misclassification rates. We achieved an average accuracy of 77.5\% with processing speed of $\approx$38~sec/GB, which is better and more than an order of magnitude faster than the previous state-of-the-art tool - Sceadan (69\% at 9 min/GB). Our tool and the corresponding dataset is open-source.
\end{abstract}

\begin{keywords}
file-type classification, memory forensics, carving, machine learning, convolutional neural network.
\end{keywords}

\section{Introduction}
\label{sec:introduction}

Memory forensics and data carving heavily rely on data/file-type identification from small chunks of memory (Fig.~\ref{fig:arch}). Many existing techniques are based on simple heuristics (like common byte markers) or hand-engineered features (unigrams~\cite{Beebe_2013,Fitzgerald_2012,Veenman_2007}, bigrams~\cite{Beebe_2013,Fitzgerald_2012}, Shannon entropy~\cite{Wang_2018,Beebe_2013,Fitzgerald_2012,Veenman_2007} or Koglomorov complexity~\cite{Wang_2018,Beebe_2013,Fitzgerald_2012,Veenman_2007}, etc. - see Section~\ref{sec:related-work} for more details). Such an approach suffers from multiple scalability issues, ranging from a limited number of supported file-types, deteriorating classification accuracy, up to slow processing speed. Feature extraction alone is responsible for a bulk of the processing time~\cite{Beebe_2013}. (Our evaluation corroborates this; see Tab.~\ref{tab:feat}.) Some studies consider boosting, and compute individual features only when necessary~\cite{li2009cocost}.

Many file carvers rely on simple heuristics and entropy thresholding to find JPEG blocks in raw disk images~\cite{7230269}. Although more sophisticated methods exist they often focus on distinguishing file-types that are very different, e.g., JPG, CSV, HTML, DOC, XLS \cite{Karresand,Veenman_2007,Beebe_2013,Calhoun_2008,Xu_2014}. Many of these file-types exhibit distinct byte frequency statistics and are also unlikely to occur in memory images from cameras and IoT devices. The complexity of the problem leads to trade-offs which can compromise the speed~\cite{Wang_2018} or accuracy~\cite{Beebe_2013,Xu_2014}. While allowing for rapid triage, reliance on simple heuristics leads to excessive false positive matches, which need to be considered and discarded at later stages. This bottleneck affects even the most recent carvers~\cite{RecoLab,7230269,uzun2015carving,durmus2019every}. 

To address these limitations, we explore modern neural networks (NNs) to dispense with the explicit feature computation. Our search for compact architectures delivering good balance between accuracy and speed resulted in \emph{FiFTy}. It is a scalable and fast file-type identification tool comprising of models based on six real-world scenarios (see Tab.~\ref{tab:scenarios}) and two common block sizes (512 and 4,096 bytes).

Our models are 1-D convolutional neural networks (CNNs) that take blocks of raw bytes as input, and embed them in a trainable latent space - akin to word embeddings that proved successful in natural language processing~\cite{mikolov2013efficient}. The resulting classifier consistently outperforms state-of-the-art models reported in the literature and can be an order of magnitude faster. We base our results on a novel dataset with 75 diverse and representative file-types (nearly twice as much as the largest prior work~\cite{Beebe_2013}). Our model achieves an average accuracy of 77.5\% (using memory blocks of 4,096 bytes) with processing speed of $\approx$38~sec/GB. The previous state-of-the art system, Sceadan~\cite{Beebe_2013}, achieved 69\% accuracy and required 9 min/GB in the same conditions (see detailed breakdown in Tab.~\ref{tab:prev}). 

While several recent studies started to explore deep learning models for the problem, their solutions have fallen short. Hiester~\cite{hiester2018file} considered fully connected, convolutional and recurrent networks, but studied only 4 very different file-types (JPEG, GIF, XML and CSV) which can be easily distinguished using simple techniques employing statistical features. The proposed models convert input byte blocks to 1-hot encoding, which leads to unnecessary dimensionality explosion. Chen et al.~\cite{Chen_2018} took a different approach, and reshaped 512-byte blocks as $64 \times 64$ grayscale images. Despite using modern CNNs and a small 16-type dataset, their approached delivered sub-par performance with accuracy of only 71\%.

Our study is significantly broader in scope. In addition to the largest reported dataset (Tab.~\ref{tab:prev}), we also study several smaller scenarios with application-specific selection of file-types. We focus on graphic and photographic data, which occur in photo carving applications. Our selection reflects the most popular files, that can be found on SD cards from modern cameras, smartphones and camera-enabled IoT devices. We also include multiple previously unstudied types like HEIC/HEIF (high efficiency image file format~\cite{hannuksela2015overview}) and numerous RAW photo varieties (including GPR from GoPro action cameras). To accommodate variations of cluster size in different file systems, we consider both 512 and 4,096-byte inputs.

To ensure fair comparison, we evaluate our approach against three strong baselines: (1) a state-of-the-art file-type classifier, \emph{Sceadan}~\cite{Beebe_2013}; (2) a fully connected neural network trained on commonly used hand-crafted global block features, denoted as \emph{NN-GF}; (3) a convolutional neural network trained on byte co-occurrences, denoted as \emph{NN-CO}. The comparison was done on the same datasets, and using the same computational resources (using GPU acceleration, when possible). The proposed approach consistently outperformed all baselines with respect to speed, overall accuracy, and individual misclassification rates.

The contributions of our work includes:
\begin{itemize}
  \item A compact neural network architecture for file-type classification with state-of-the-art performance and runtime; we also explore the hyperparameter space using the Tree-structured Parzen Estimator (TPE)~\cite{bergstra2011algorithms}.

  \item Detailed evaluation of the proposed approach against three strong baselines, including a state-of-the-art tool based on support vector machines~\cite{Beebe_2013}, and modern neural networks trained on hand-crafted features.
  
  \item Detailed analysis and discussion of several classification scenarios relevant to photo carving applications.
  
  \item A ready-to-use, open-source implementation (Python 3) available at \url{https://github.com/mittalgovind/fifty} and via \texttt{pip3 install fifty}.
  
  \item A novel data-set (75 popular file-types) publicly available at \url{http://dx.doi.org/10.21227/kfxw-8084}.
\end{itemize}
  
The paper is organized as follows. Section~\ref{sec:related-work} reviews related work. In Section \ref{sec:proposed-model}, we describe the proposed model, discuss key design decisions and hyper-parameter selection. We report the results of experimental evaluation in Section~\ref{sec:experiments} and conclude in Section~\ref{sec:conclusions}.

\section{Related Work}
\label{sec:related-work}

\emph{Data carving} is a technique used to identify and extract specific file-types from memory blocks, without assuming explicit access to file-system metadata. The simplest carving methods rely on starting and ending signatures (e.g., magic numbers). However, they are inadequate when dealing with heavily fragmented file-systems or in the presence of address space randomization (often used to alleviate memory cell wear in modern SSDs). Similar techniques are also applicable to \emph{memory forensics}, which involves analyzing random access memory (RAM) dumps. In such cases, it is helpful to use a file-type classifier that can blindly identify the data-type of small memory blocks.

Most of file-type classifiers relied on hand-crafted heuristic features, and were limited to very small datasets. Only recently researchers began exploring (deep) neural networks, but the reported solutions were still limited in both scope and performance. Tab.~\ref{tab:prev} summarizes the previous approaches, their evaluation scenarios, and the reported performance. The table is divided into two sections: the top 4 rows summarize results of our evaluation of \emph{FiFTy} and three baseline methods (\emph{Sceadan, NN-GF, NN-CO}) in the same conditions (all 75 file-types; 4,096-byte blocks; same hardware); the remaining rows summarize results, as reported in their original publications.

\subsection{Hand-crafted Features}

Karresand et al.~\cite{Karresand} developed \emph{Oscar}, a classifier based on two features: rate of change (RoC) of consecutive bytes, and binary frequency distribution (BFD). They defined a 1-norm metric as the distance of these features from pre-computed centroids. This approach worked well for JPEG photos due to \emph{byte-stuffing}~\cite{wallace1992jpeg}, a technique involving insertion of \texttt{0x00} after \texttt{0xFF}. The RoC distribution captured this feature easily and thus lead to near-perfect detection of JPEG files (99.2\%) but with high false positive rates for other high-entropy filetypes, e.g., Windows executables and ZIP. 

Veenman~\cite{Veenman_2007} trained a supervised classifier on a dataset described by 1-grams, entropy, and Kolmogorov complexity. They employed a two-stage strategy with multi-class and binary classification (one-against-all) using the Fischer linear discriminant (FLD) aiming to maximize inter-class distances. While this method yielded high accuracy for JPEG files (98\%), it averaged to only 45\% on a dataset with 11 filetypes. Calhoun et al.~\cite{Calhoun_2008} also used the FLD, but included additional features, e.g., longest common strings and subsequences. While they report good results, and consider various feature combinations, the study was limited to only 4 filetypes (JPEG, PDF, GIF and BMP).

Fitzgerald et al.~\cite{Fitzgerald_2012} used unigrams and bigrams along with entropy and complexity measures to train a support vector machine (SVM) based classifier. They used the bag-of-words model (here, bag-of-bytes) to classify text documents (file fragment). This method worked well for text formats but failed for all high entropy filetypes.

Beebe et al.~\cite{Beebe_2013} developed \emph{Sceadan}, a tool based on various statistical features, including unigrams, bigrams, entropy, longest streak, etc. The features were originally extracted from 512-byte blocks, and organized into four sets: unigram, bigram, all other global features and a subset of global features. These feature sets were then used for classification using SVMs with linear and radial basis function (RBF) kernels. The authors concluded that increasing the number of types of features had a negative effect on the classification accuracy, and that unigrams \& bigrams worked best as they are. Until now, \emph{Sceadan} remains the state-of-the-art open source tool and the corresponding paper is the most comprehensive study to date - it covered 38 diverse filetypes and reported accuracy of 73.8\%. We used \emph{Sceadan} as an important baseline in our experiments. In Tab.~\ref{tab:prev}, we collect both the results reported by the authors, as well as the results from our evaluation (with both 512 and 4096-byte blocks). More information about the experiment is available in Section ~\ref{sec:baseline}.

Xu et al.~\cite{Xu_2014} converted 512-byte blocks into $32 \times 32$~px grayscale images and extracted a 512-D GIST image descriptors based on Gabor filters. Finally, they  experimented with various classification algorithms, (e.g., Naive-Bayes, SVMs, k-nearest neighbours, SMO, J48) and concluded that KNN performed best both in terms of speed and accuracy.

Zheng et al.~\cite{Zheng_2015} also relied on SVMs but employed only two features (binary frequency distribution and Shannon entropy). They explicitly differentiate between \emph{file-types} (that could contain other embedded files, e.g., PDF and DOC) and \emph{data-types} (that are often embedded within other files, e.g., JPEG photos), but ultimately deal only with filetypes without any embedded foreign objects.

Beebe et al.~\cite{Beebe_2016} experimented with clustering of 52 filetypes, and proposed a two-layer hierarchical structure where filetypes are first grouped into 6 classes and classifiers are trained separately for each class. The authors used the same features as all other studies, including unigrams, bigrams and global statistical features (mean, entropy, complexity, etc.). The paper focuses on exploratory analysis based on K-means clustering, and reports file-level accuracy of 74.1\%. The utilized classification strategy is not reported.

\subsection{Automatic Feature Extraction}

Wang et al.~\cite{Wang_2018} used sparse coding for automatic feature extraction. They computed higher-order $N$-grams ($N$ = 4, 8, 16, 32, 64) and stored them in dictionaries. The dictionaries are subsequently averaged and concatenated with bigram and unigram frequencies. Similarly to other studies, feature extraction takes most of the processing time and makes the approach infeasible for high-volume applications. The reported accuracy was also sub-par, reaching on average 61.3\% for 18 filetypes.

Hiester~\cite{hiester2018file} considered fully connected, convolutional and recurrent neural networks with raw bytes fed directly as model inputs (1-hot encoded). The study included only 4 very different filetypes (JPEG, GIF, XML and CSV) which are easy to distinguish with existing techniques. The adopted 1-hot encoding contributes to excessive dimensionality and slow training. Only the recurrent network achieved satisfactory accuracy, and did so at excessive computational effort. The presented approach is not scalable to real-world applications.

\begin{sidewaystable*}
  \caption{\label{tab:prev} Quantitative review of related work in reverse chronological order; Reported metrics correspond to: our experiments on the same dataset (75 file-types; 512 and 4,096-byte inputs) and hardware (top 4 rows); and results reported in original publications (all remaining rows)}
  \resizebox{\textwidth}{!}{\begin{tabular}{ccllllllllllc}
 \toprule
 \textbf{Year} & \textbf{Classifier} & \textbf{Features} & \parbox[t]{0.05\textwidth}{\textbf{Feature dim.}} & \parbox[t]{0.03\textwidth}{\textbf{\#File types}} & \parbox[t]{0.03\textwidth}{\textbf{Block size}} & \textbf{ Accuracy} & \parbox[t]{0.05\textwidth}{\textbf{JPEG Acc.}} &  \parbox[t]{0.07\textwidth}{\textbf{Speed {[}ms/block{]}}$^\dagger$} & \parbox[t]{0.08\textwidth}{\textbf{Speed {[}min/GB{]}}$^\dagger$} & \parbox[t]{0.09\textwidth}{\textbf{Training Time}$^\ddagger$} & \textbf{Remarks} & \textbf{Ref.}\\ 
 \midrule

  2019 &
  \multirow{2}{*}{CNN}&
  \multirow{2}{*}{\parbox[t]{0.15\textwidth}{raw bytes}} &
   n/a &
  75$^1$ &
  512 &
  65.6 &
  83.5 &
  0.04&
  1.50&
  7.51 hours& 
\multirow{2}{*}{\parbox[t]{0.20\textwidth}{\emph{FiFTy}}}&
  \multirow{2}{*}{Self} \\ 
 
  &
  &
  &
  n/a &
  75$^1$ &
  4096 &
  77.5&
  86.3&
  0.15&
  0.64&
  12.53 hours&
  & 
  \\ 
  \midrule
   2019 &
  \multirow{2}{*}{CNN}&
  \multirow{2}{*}{\parbox[t]{0.15\textwidth}{co-occurence matrix}} &
  \multirow{2}{*}{(128, 128)} &
  75$^1$ &
  512 &
  64.4 &
  77.0 &
  0.84 &
  29.36 &
  8.67 hours&
  \multirow{2}{*}{\parbox[t]{0.20\textwidth}{\emph{NN-CO}}} &
  \multirow{2}{*}{Self} \\ 
 
  &
  &
  &
  &
  75$^1$ &
  4096 &
  75.3 &
  92.5 &
  1.05 &
  4.59 &
  9.04 hours&
  & 
  \\ 
  \midrule
  2019 &
  \multirow{2}{*}{NN}&
  \multirow{2}{*}{\parbox[t]{0.15\textwidth}{global stat. features}} &
  \multirow{2}{*}{14} &
  75$^1$ &
  512 &
 45.4 &
  53.2 &
  0.61&
  21.32&
  2.88 hours&
  \multirow{2}{*}{\parbox[t]{0.20\textwidth}{\emph{NN-GF}}} &
  \multirow{2}{*}{Self} \\ 
 
  &
  &
  &
  &
  75$^1$ &
  4096 &
  46.8 &
  48.6 &
  2.67&
  11.67 &
  10.53 hours&
  & 
  \\ 
  \midrule

  2019 &
  SVM & 
  \multirow{2}{*}{\parbox[t]{0.15\textwidth}{statistical: unigram, bigram, entropy, etc}} &
   \multirow{2}{*}{65792} &
  75$^1$&
  512 &
  \parbox[t]{0.01\textwidth}{{57.3}} &
  \parbox[t]{0.01\textwidth}{81.5} &
  1.48 &
  51.7 &
  44 hours &
  \multirow{2}{*}{\parbox[t]{0.20\textwidth}{\emph{Sceadan} \cite{Beebe_2013}; our evaluation}} &
  \multirow{2}{*}{{Self}}
   \\ 
  
  &
  &
  &
  &
  75$^1$ &
  4096 &
  69.0 &
  91.8 &
  2.06 &
  9.0 &
  10 days* &
   & 
   \\ 
   \midrule \midrule
  2018 &
  CNN &
  \multirow{2}{*}{\parbox[t]{0.15\textwidth}{1-hot enc. bytes}}&
  \multirow{2}{*}{8192} &
  4$^2$ &
  512 &
  73.4&
  \parbox[t]{0.01\textwidth}{96.5} &
  n/a &
  n/a &
  7 hours &
  \multirow{2}{*}{\parbox[t]{0.20\textwidth}{Few \& very different file types}} & 
  \multirow{2}{*}{ \cite{hiester2018file}}
   \\ 

  &
  RNN &
  &
  &
  4$^2$ &
  512 &
  98.04 &
  \parbox[t]{0.01\textwidth}{96.5} &
  n/a &
  n/a &
  135 hours &
  & \\ 
  \midrule
  2018 &
  CNN &
  \parbox[t]{0.15\textwidth}{image pixels}&
  (64, 64) &
  16$^2$ &
  4096 &
  \parbox[t]{0.01\textwidth}{70.9} &
  \parbox[t]{0.01\textwidth}{15 } &
  n/a &
  n/a &
  n/a &
  \parbox[t]{0.20\textwidth}{}&
   \cite{Chen_2018}\\ 
  \midrule
  2018 &
  SVM &
  \parbox[t]{0.15\textwidth}{dictionary of N-Grams}&
  \parbox[t]{0.05\textwidth}{1024 + 1296 + 1444} &
  18$^2$ &
  512 &
  \parbox[t]{0.01\textwidth}{61.3} &
  \parbox[t]{0.01\textwidth}{65.0} &
  n/a &
  n/a &
  \parbox[t]{0.05\textwidth}{235 ms/block} &
  \parbox[t]{0.20\textwidth}{Feature extraction is the bottleneck}
  &  
   \cite{Wang_2018} \\ 
  \midrule
  2016 &
  K-means &
  \parbox[t]{0.20\textwidth}{unigrams, bigrams, and global statistics (mean, entropy, complexity, etc.)}
  &
  n/a &
 \parbox[t]{0.07\textwidth}{52$^{2,3}$ \\ (6 classes)} & 
  512 &
  \parbox[t]{0.01\textwidth}{74.0} &
   n/a &
   n/a &
   n/a &
   n/a &
   \parbox[t]{0.20\textwidth}{Focus on exploratory heirarchical clustering} &
    \cite{Beebe_2016}\\ 
 \midrule
 
  2015 &
  SVM &
  \parbox[t]{0.15\textwidth}{BFD, entropy}&
  n/a &
  12$^2$ &
  512 &
  \parbox[t]{0.01\textwidth}{67.4} &
  \parbox[t]{0.01\textwidth}{69.2} &
  n/a &
  n/a &
  n/a &
  \parbox[t]{0.20\textwidth}{Rudimentary analysis} &
   \cite{Zheng_2015}\\ 
  \midrule
  2014 &
  KNN  &
  image pixels &
  \parbox[t]{0.06\textwidth}{512 ($32 \times 32$)} &
  29$^2$ &
  1024 &
  \parbox[t]{0.01\textwidth}{39.7} &
  \parbox[t]{0.01\textwidth}{22 } &
  n/a &
  n/a &
 \parbox[t]{0.05\textwidth}{0.001 ms/block} &
 \parbox[t]{0.20\textwidth}{Includes various classifiers \& grid search}
  &  \cite{Xu_2014}\\ 
  \midrule
  2013 &
  SVM &
  \parbox[t]{0.15\textwidth}{statistical: unigram, bigram, entropy, etc} &
  65792 &
  38$^{2,3}$ &
  512 &
  \parbox[t]{0.01\textwidth}{73.8} &
  \parbox[t]{0.01\textwidth}{81 } &
  0.30&
  10.6&
  n/a &
  \parbox[t]{0.20\textwidth}{\emph{Sceadan}; as reported by authors} &
   \cite{Beebe_2013} \\ 
 
  \midrule
 
  2012 &
  SVM &
  \parbox[t]{0.17\textwidth}{unigram, bigram, entropy, Hamming weight, complexity, longest cont. streak}&
   \parbox[t]{0.05\textwidth}{$256 +256^2 + 6$} &
  24$^2$ &
  512 &
  49.1 &
  \parbox[t]{0.01\textwidth}{17.4} &
  n/a &
  n/a &
  n/a &
  \parbox[t]{0.20\textwidth}{Poor results for all high-entropy file types (incl. JPG)}&
   \cite{Fitzgerald_2012}\\
  \midrule
  2008 &
  FLD &
  \multirow{2}{*}{\parbox[t]{0.18\textwidth}{longest common substrings, other statistical measures}}  &
  21 &
  4$^3$ &
  512 &
  \parbox[t]{0.01\textwidth}{86 } &
  \parbox[t]{0.01\textwidth}{83.6} &
  n/a &
  n/a &
  n/a &
  \multirow{2}{*}{\parbox[t]{0.20\textwidth}{Pair-wise analysis}} &
   \cite{Calhoun_2008} \\ 
 
   &
  &
  &
  - &
  4$^3$ &
  896 &
  \parbox[t]{0.01\textwidth}{88.3} &
  \parbox[t]{0.01\textwidth}{92} &
  n/a &
  n/a &
  n/a &
  &
   \\ 
 
  \midrule
 
  2007 &
  FLD &
  \parbox[t]{0.15\textwidth}{1-gram, entropy, Kolmogorov complexity}&
  n/a &
  11$^3$ &
  4096 &
  \parbox[t]{0.01\textwidth}{45 } &
  98 &
  n/a &
  n/a &
  n/a &
  & 
   \cite{Veenman_2007}\\ 
  \midrule
 
  2006 &
  VQ &
  \parbox[t]{0.15\textwidth}{rate of change and binary freq. distribution}&
  n/a &
  4$^3$ &
  512 &
  n/a &
  \parbox[t]{0.01\textwidth}{99.2} &
  0.02 &
  0.6 &
  n/a &
  \parbox[t]{0.20\textwidth}{JPEG-specific method (byte markers); high FPR for other types} &
   \cite{Karresand} \\
 
  \bottomrule
  \multicolumn{13}{l}{Dataset: $^1$ - the proposed FFT-75 dataset\cite{kfxw-8084-19} ; $^2$ - a public dataset GovDocs1~\cite{govdocs1} ; $^3$ - Private dataset; * ref. section~\ref{sec:eval_res}; $^\dagger$ -  Computed on Tesla V100 (Top 4 rows) ; $^\ddagger$ - Computed on Tesla P40 (Top 4)}\\ 
 
 \end{tabular}}
\end{sidewaystable*}

Chen et al.~\cite{Chen_2018} took an approach similar to Xu et al.~\cite{Xu_2014} by converting 512-byte fragments into $64 \times 64$ images. Despite using a CNN architecture, the model delivered sub-par performance with accuracy of only 71\% among 16 filetypes. Moreover, the network failed to classify even very distinct types like HTML or plain text files.

\section{Proposed Method}
\label{sec:proposed-model}

This section discusses our motivations for the approach, use-cases, description of the general architecture of models included in \emph{FiFTy} and the dataset used.

\subsection{Formal Problem Statement}

We address the problem of inferring the file-type of a block of bytes in the absence of any side information, i.e., we aim to obtain a function $f$ that maps a single block $B$ of $n$ bytes to its file-type label $T$: 
\begin{equation}
 f: B \in \mathbb{Z}_{255}^n \rightarrow T \in \{\mathrm{JPG, PPT, ..., PNG}\}
\end{equation}
\noindent where $\mathbb{Z}_{255} = [\texttt{0, \ldots, 255}]$. We will model function $f$ as a neural network, which allows for jointly learning a suitable feature representation, and the final classifier.

We focus on the more common \emph{file-type classification} and not on \emph{data-type classification}. The latter is more granular, and involves complex combinations where one data type might be embedded into another (e.g., images embedded into PDF documents, base64-encoded images embedded into HTML, or sections of structured data within human-readable text). This leads to prohibitive complexity of training data preparation, more complex carving logic, and still does not solve the necessity to handle heterogeneous blocks with multiple data types (especially for larger blocks). Hence, \emph{file-type classification} is often the best choice in practice (cf. Section \ref{sec:related-work}).

\subsection{Design Objectives}

In this work, we focus on three objectives -- \emph{speed, accuracy} and \emph{generality} -- to make design choices. We aim to meet these objectives by relying on modern neural networks, which can automatically learn effective feature representations from raw bytes and dispense with explicit feature computation - a major bottleneck in previous solutions~\cite{Beebe_2013} (Tab.~\ref{tab:feat}). Moreover, NN models can take advantage of advanced accelerators (GPUs, TPUs, VPUs) and can be easily deployed in desktop, server, mobile and edge environments~\cite{intelMovidius}. We used \emph{Keras} which abstracts model definition and allows to exchange computational backends~\cite{chollet2015keras}. In our experiments, we used Python 3.6 and the Tensorflow (1.5.0) backend. All experiments were run on a server with 4 Intel Xeon CPUs (E5-2680 v4 @ 2.40GHz) and a single Tesla V100 GPU. Our models achieve nearly identical runtime regardless of the number of output classes (Tab.~\ref{tab:scenarios}).

To facilitate representation learning, we collected a novel dataset with 75 filetypes including: 6 bitmap photo formats, 11 RAW photo formats, 7 video formats, 7 audio formats, and many others. Hence, not only is it the largest reported dataset, but also the only one uniquely suited for data carving applications. However, to make our discussion as general as possible, we consider 6 application scenarios with different levels of granularity (see Section~\ref{sec:scenarios}). 

\begin{figure}[]
  \centering
  \includegraphics[width=0.7\columnwidth]{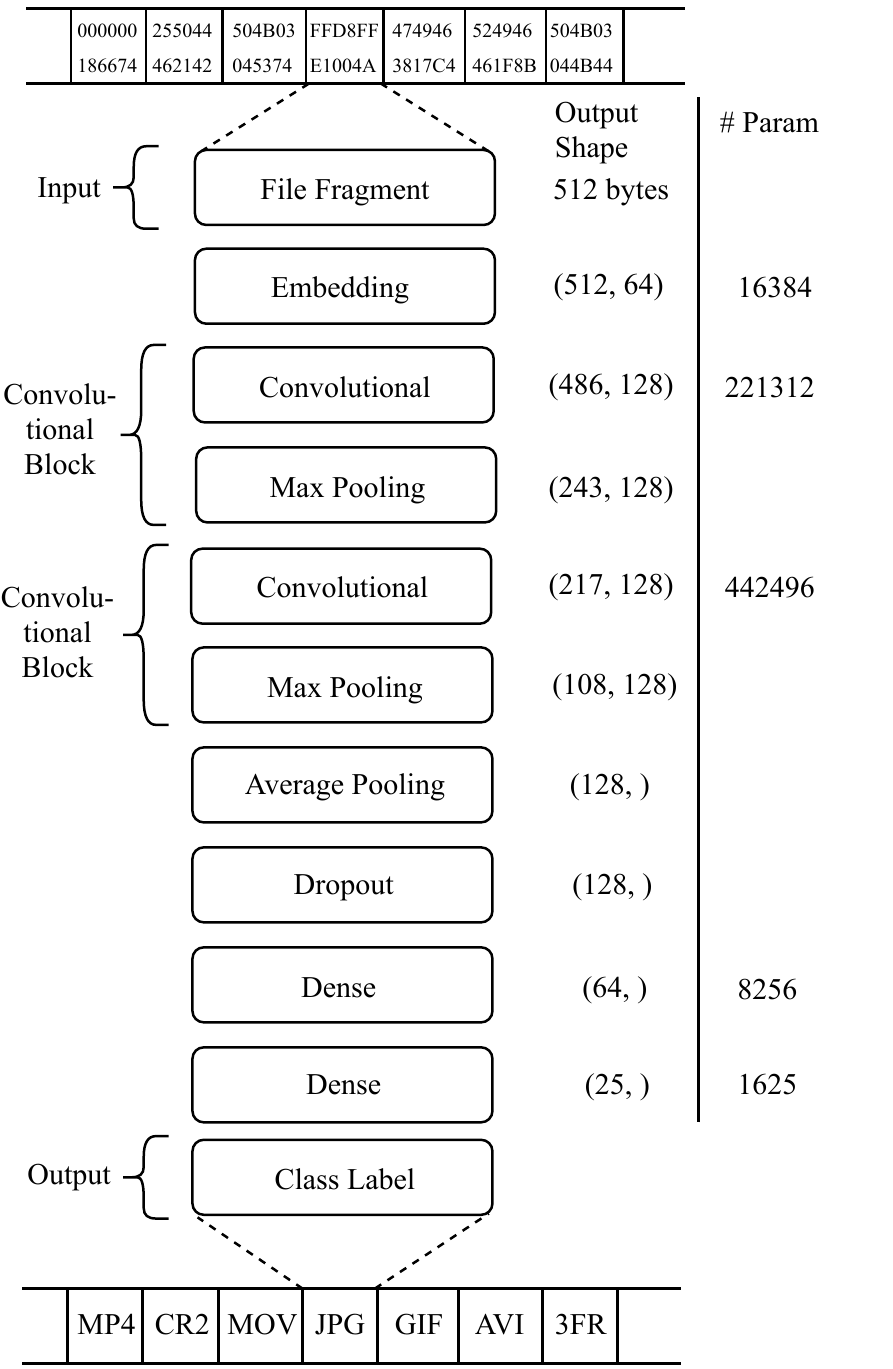}
  \caption{Illustration of the proposed network architecture; the presented model corresponds to hyper-parameters optimized for scenario \#3 w. 512-byte inputs; the reported numbers indicate how dimensionality changes with successive processing steps, and how parametrization is allocated throughout the model.}
  \label{fig:arch}
\end{figure}

\subsection{General Model Architecture}
\label{sec:models}

We feed raw byte blocks as the input to our models. Instead of one-hot encoding, we use a trainable \emph{embedding layer} in a manner similar to natural language processing (NLP)~\cite{mikolov2013efficient}. In our work, we treat file-fragments as a sequence of bytes and intend to capture characteristic byte transitions patterns, inherent to many file-types.

Embedding layers are commonly used for learning a fixed-length real-valued representation of variable-length discrete words, while maintaining their linguistic context~\cite{mikolov2013efficient}. In NLP, word embeddings can learn relationships between concepts, and allow for meaningful arithmetics. In our work, we use the embedding layer mainly for their efficiency in converting sparse binary features (1-hot encoded words in the dictionary) into compact real-valued representations. This allowed us to reduce a single byte (256 binary features in 1-hot encoding) into 16-64 real-valued features, which are more meaningful for machine learning models. The following layers will further reduce this dimensionality by pooling spatially-adjacent features of intermediate representations.

Our architecture is 1-dimensional and relies on convolutions, followed by max-pooling to gradually reduce the size of the problem. We refer to these units as \emph{convolutional blocks} and vary their number as a tunable hyper-parameter. Our final vectorized representation is obtained by average pooling with dropout. In the last step, two fully connected layers map the obtained feature vectors into class probabilities. All hidden layers are activated by \emph{Leaky Rectified Linear Units (LeakyReLU)} with $\alpha = 0.3$.

The overall architecture is shown in Fig.~\ref{fig:arch}. We used the same general architecture for all tested scenarios (Section~\ref{sec:scenarios}; Tab.~\ref{tab:scenarios}) and only vary hyper-parameters (Section~\ref{sec:hyperparameters}). The model shown in the figure corresponds to hyper-parameters optimized for scenario \#3 with block size of 512 bytes. The reported numbers indicate how dimensionality changes with successive processing steps, and how parametrization is allocated throughout the model.

We also experimented with recurrent architectures using GRU units placed after the last convolutional layer. While initially this seemed to be more accurate, careful hyper-parameter tuning eventually reduced performance gap to 1\%, which no longer justified nearly 40$\times$ increase in processing time. While probably better performance could be obtained, longer processing time makes it difficult to explore the hyper-parameter space. As a result, we currently believe that carefully tuned 1-D convolutional models are the best choice with respect to the accuracy-speed trade-off. 

\subsection{Design Space and Hyper-parameter Optimization}
\label{sec:hyperparameters}

We use the same general network architecture described in Section~\ref{sec:models} for all our experiments, and only tune selected hyper-parameters to adjust the model to the scenario at hand. The scenarios differ in the number of output classes and reflect various application needs, ranging from general-purpose to increasingly specialized photo carving (Section~\ref{sec:scenarios}).

\subsubsection{Hyper-parameter Definition}

\begin{figure}[t]
  \centering
  \subfloat[][After initial 20 random samples TPE selects 64 as the next value.]{\includegraphics[width=\columnwidth]{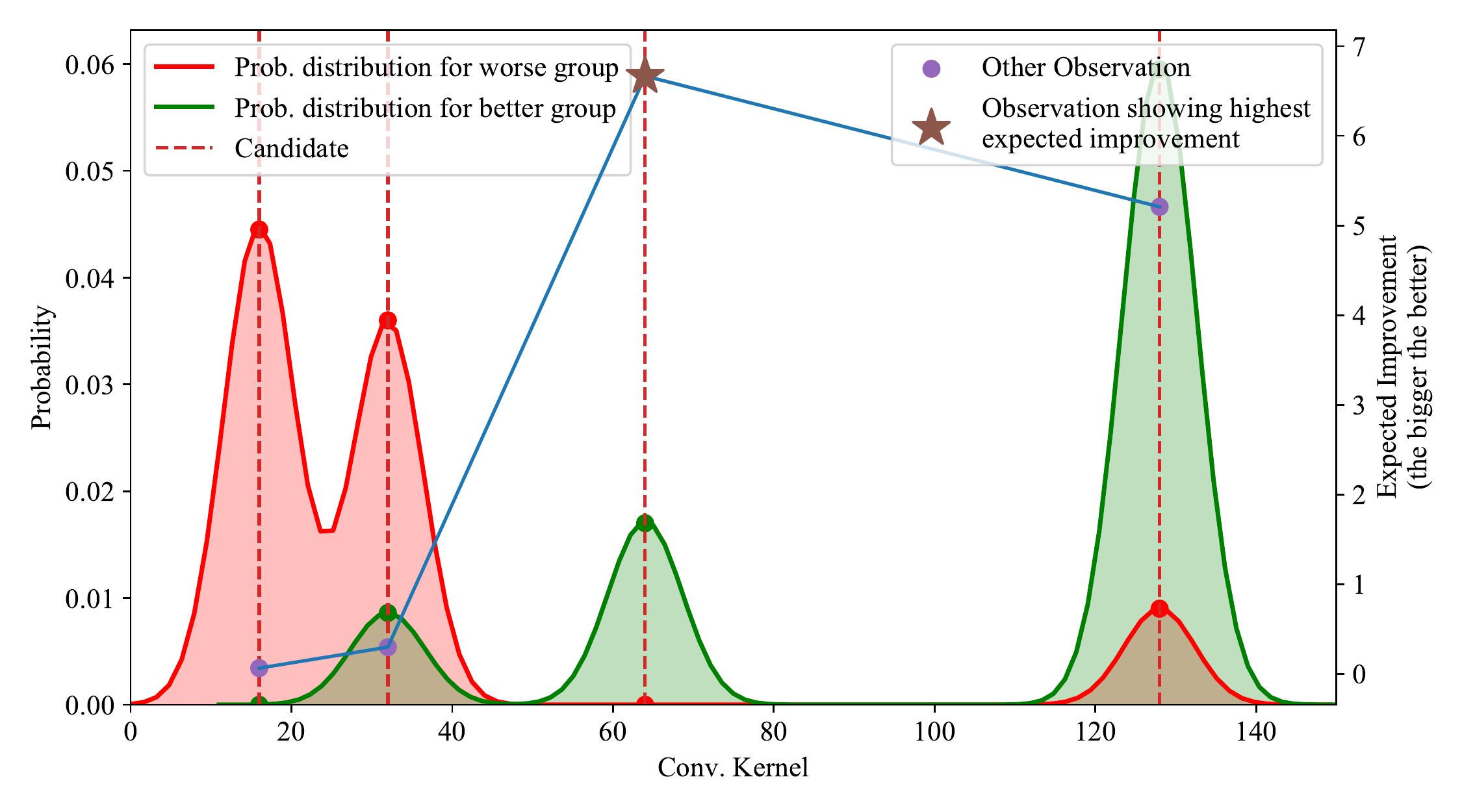}} \\ 
  \subfloat[][After 200 samples, 128 seems to be the optimal value.]{\includegraphics[width=\columnwidth]{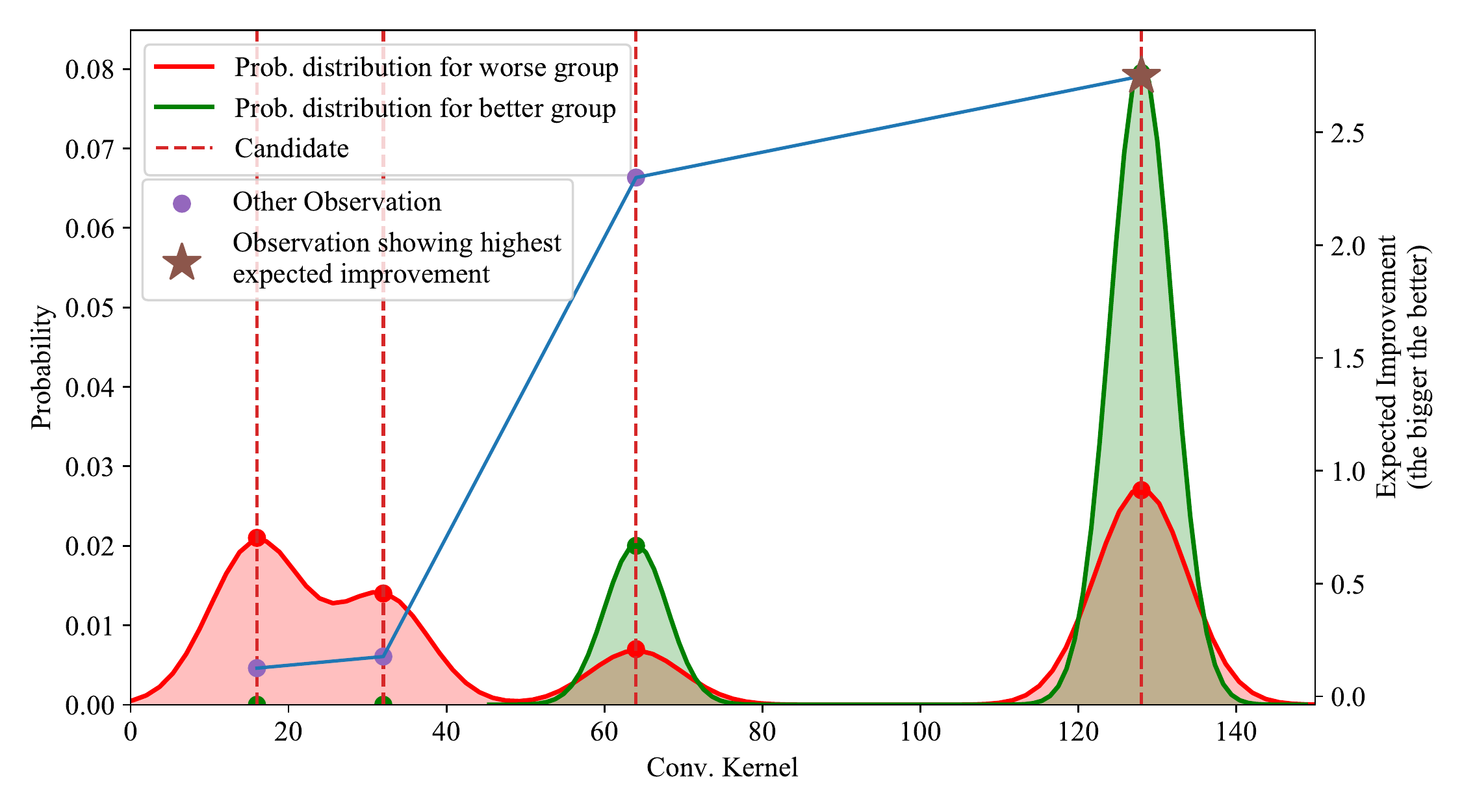}}
  \caption{Hyper-parameter value selection (here conv. kernel size) using TPE: probability distributions for top performers (red) and others (green) and the \emph{expected improvement} (blue line).}
  \label{fig:tpe}
\end{figure}

After hand tuning the general architecture, we defined 6 hyper-parameters which will be optimized automatically using the Tree-structured Parzen Estimator (TPE)~\cite{bergstra2011algorithms}. For each dimension, we chose a few possible values spanning a wide range of reasonable choices. 
  
\begin{itemize}
  \item \emph{\# dense units:} the number of units in the hidden dense layer; $\{16, 32, 64, 128, 256\}$;

  \item \emph{embedding size:} dimensionality of the trainable real-valued embedding space; $\{16, 32, 48, 64\}$;
  
  \item \emph{conv. kernel:} the size of the convolution kernel (same for all layers); $\{16, 32 , 64, 128\}$;
  
  \item \emph{conv. stride:} convolution stride (same for all layers): $\{3, 11 , 18, 27, 35\}$;
  
  \item \emph{\# conv. blocks:} the number of convolutional blocks (conv +  max-pooling); $\{1, 2, 3\}$;
  
  \item \emph{max-pooling size:} max pooling window size; $\{2, 4, 6, 8\}$.
\end{itemize}

\noindent In a separate, prior experiment, we established that adoption of \emph{dropout} is beneficial before the hidden dense layer (with dropping probability of 0.1).

\begin{figure}[t]
  \centering
  \includegraphics[width=0.85\columnwidth]{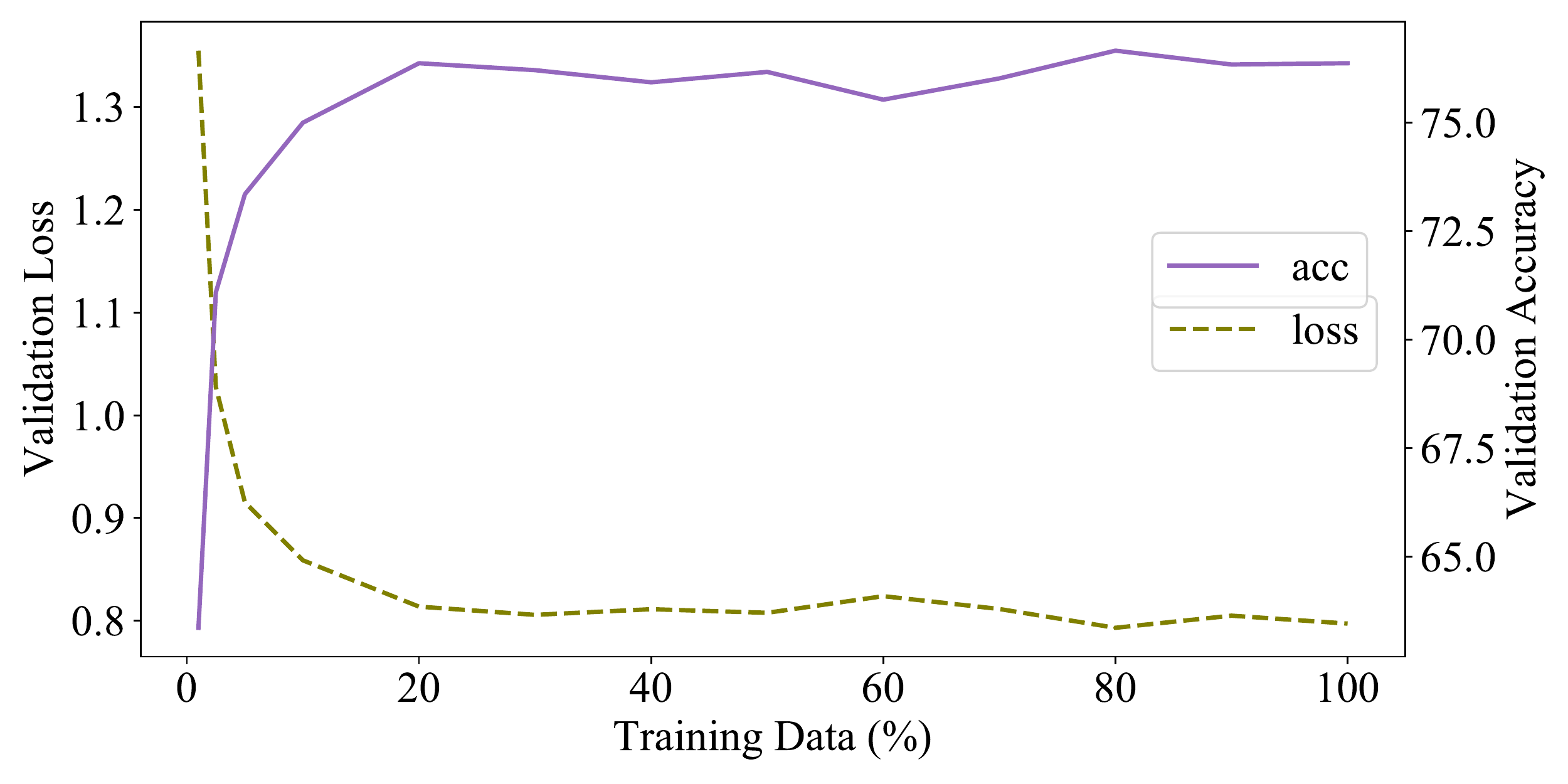}
  \caption{Impact of training data size on validation accuracy and loss (scenario \#1 with 4,096-byte blocks).}
  \label{fig:data-size-impact}
\end{figure}

In Fig.~\ref{fig:data-size-impact}, we graph validation loss and accuracy achieved by the same model versus the amount of data used for training. We observe that even using less than 20\% data gave us proportional results that could be used to compare the models. Therefore, to speed up the hyper-parameter space exploration, we used 10\% of the training data and 40\% of the validation data instead of using the full set of data everytime.

We sampled network configurations using the TPE (we used the implementation available in the \emph{hyperopt} package~\cite{bergstra2013making}). In total, we collected 225 models for each of the 12 scenarios (6 base scenarios with 512 and 4,096-byte inputs, respectively). The first 20 configurations are fully random, and the remaining are sampled to maximize model performance. Most scenarios completed within a day, except for the largest two (scenario \#1 with 75-classes for both 512 and 4096-byte inputs) that took up to 36 hours. Finally, we retrained the best candidate models on the entire dataset, and used a separate hold-out test set for the final evaluation.

\subsubsection{Tree-structured Parzen Estimator}

\begin{figure*}[t!!]
  \centering
  \includegraphics[width=\textwidth]{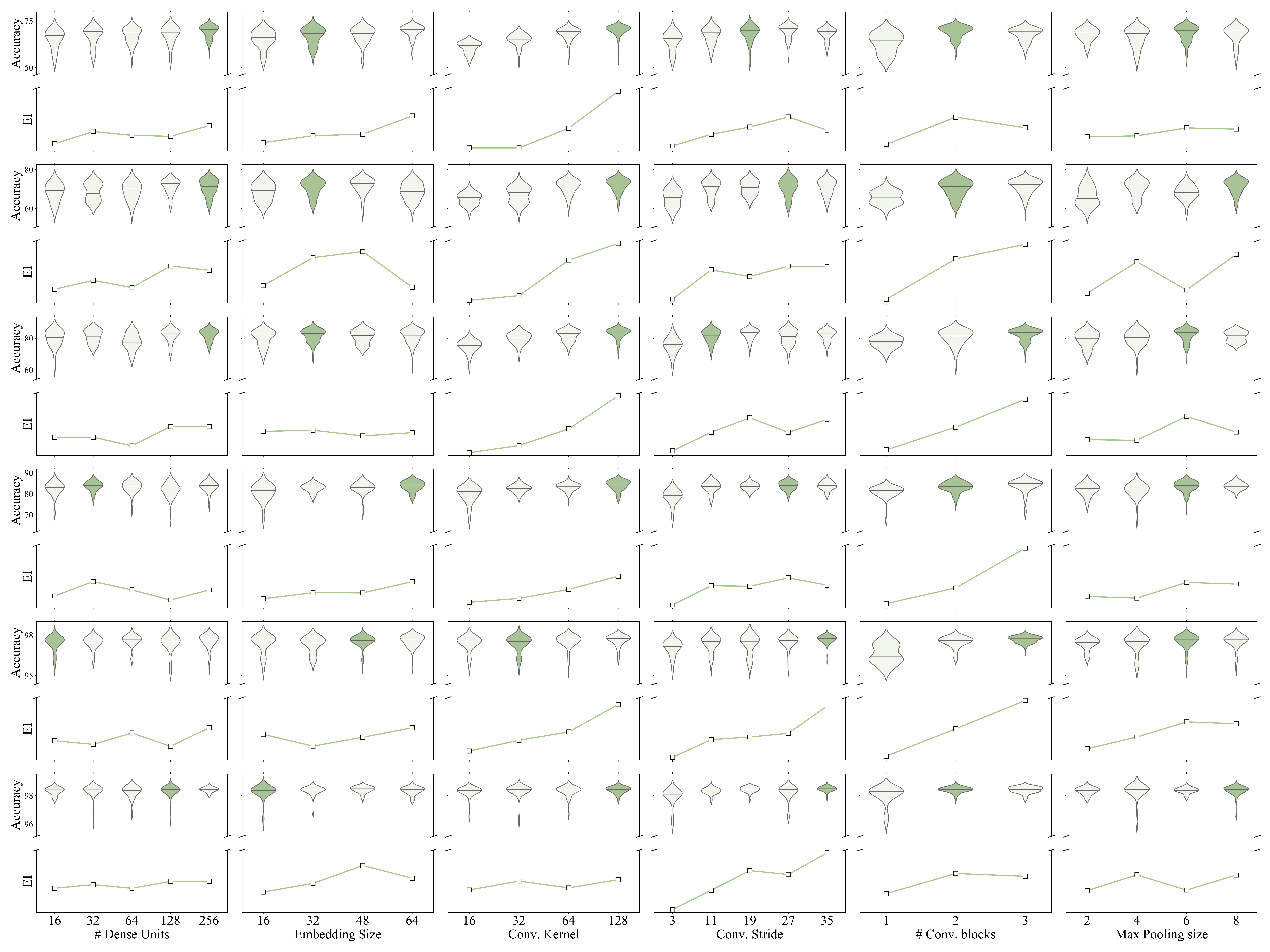}
  \caption{Impact of hyper-parameter variation on the validation accuracy for all scenarios with 4096-byte inputs: (top) violin plots showing distribution of accuracy for sampled networks - dark color corrosponds to the TPE chosen value; (bottom) \emph{expected improvement} estimates from TPE.}
  \label{fig:hyper_4k}
\end{figure*}

TPE is a sequential model-based optimization (SMBO) algorithm~\cite{ Hutter:2011:SMO:2177360.2177404} proposed by Bergstra et al.~\cite{bergstra2011algorithms}, which models the probability that given a scoring metric $L$, a hyperparameter value $\theta$ leads to a better performance, denoted as $p(\theta|L)$. The algorithm samples subsequent values for evaluation based on historical performances.

The algorithm can be summarized as follows:

 \begin{enumerate}
  \item Initialize the model with $N_\alpha$ random samples from the hyper-parameter space $S$.
  \item Divide the sampled models into two groups based on a threshold on the performance quantile. 
  \item Estimate the probability density functions for top performers ($p_1$) and the rest ($p_2$).
  \item The next sample $\hat{\theta}$ is determined using the \emph{expected improvement}:  
    \begin{equation}
      \hat{\theta} = \underset{\theta \in \Theta}{\text{argmax }} EI (\theta) = \underset{\theta \in \Theta}{\text{argmax }} \frac{p_1(\theta)}{p_2(\theta)},
    \end{equation}
  where $p_1(\theta)$ and $p_2(\theta)$ are the group membership probabilities defined above. 
  \item Evaluate the network defined by $\hat{\theta}$ and repeat steps 2 to 5 until $N$ samples are collected.
\end{enumerate}

Fig.~\ref{fig:tpe} shows the modeled probabilities and the expected improvement (EI) for an example hyper-parameter from our design space (conv. kernel). Fig.~\ref{fig:tpe}(a) illustrates the initialization step of TPE. It collected 20 random samples and divided it into two equal groups using a quantile (here, 0.5). The EI is calculated and the value of 64 maximizes it. Fig.~\ref{fig:tpe}(b) illustrates the distributions after 200 samples have been collected, showing that the worse sample distribution is moving towards the better sample distribution (here, towards higher kernel size).

\subsubsection{Final Models and Key Observations}

\begin{table*}[t]
  \caption{Model architectures obtained with TPE for each scenario and block size.}
  \label{tab:models}
  \centering
  \resizebox{\textwidth}{!}{\begin{tabular}{C{1.1cm}L{14.6cm}R{1.5cm}}
  \toprule
  \textbf{Scenario} & \multicolumn{1}{c}{\textbf{Model Description}} & \textbf{\# Params}  \\ \midrule
  & \multicolumn{2}{c}{\textbf{\emph{FiFTy} models for 512-byte blocks}} \\ \midrule 
  \#1 & E (64) - C1D (128, 27) - MP (4) - AP - D (0.1) - F (256) - F (75) & 289,995 \\ 
  \#2 & E (48) - C1D (128, 11) - MP (4) - C1D (128, 11) -  MP (4) - AP - D (0.1) - F (64) - F (11) & 269,323 \\ 
  \#3 & E (64) - C1D (128, 27) - MP (2) - C1D (128, 27) -  MP (2) - AP - D (0.1) - F (64) - F (25) & 690,073 \\ 
  \#4 & E (48) - C1D (128, 19) - MP (4) - C1D (128, 19) -  MP (4) - AP - D (0.1) - F (256) - F (5) & 474,885 \\ 
  \#5 & E (64) - C1D (128, 35) - MP (8) - AP - D (0.1) - F (256) - F (2) & 336,770 \\ 
  \#6 & E (32) - C1D (128, 11) - MP (6) - C1D (128, 11) -  MP (6) - AP - D (0.1) - F (64) - F (2) & 242,114 \\ \midrule 
  & \multicolumn{2}{c}{\textbf{\emph{FiFTy} models for 4,096-byte blocks}} \\ \midrule
  \#1 & E (32) - C1D (128, 19) - MP (4) - C1D (128, 19) - MP (4) -  AP - D (0.1) - F (256) - F (75) & 449,867 \\  
  \#2 & E (32) - C1D (128, 27) - MP (8) - C1D (128, 27) -  MP (8) - AP - D (0.1) - F (256) - F (11) & 597,259 \\ 
 \#3 & E (32) - C1D (128, 11) - MP (6) - C1D (128, 11) -  MP (6) - C1D (128, 11) - MP (6) - AP - D (0.1) - F (256) - F  (25)  & 453,529 \\ 
  \#4 & E (64) - C1D (128, 27) - MP (6) - C1D (128, 27) -  MP (6) - AP - D (0.1) - F (32) - F (5) & 684,485 \\ 
 \#5 & E (48) - C1D (32, 35) - MP (6) - C1D (32, 35) -  MP (6) - C1D (32, 35) - MP (6) - AP - D (0.1) - F (16) - F (2) &  138,386 \\ 
  \#6 & E (16) - C1D (128, 35)- MP (8) - C1D (128, 35) -  MP (8) - AP - D (0.1) - F (128) - F (2) & 666,242 \\ 
  \toprule
   \multicolumn{1}{c}{\textbf{Baseline}} & \multicolumn{2}{c}{\textbf{Baseline neural network models (scenario \#1 for 512 and 4,096-byte blocks)}} \\  \midrule
 
  NN-CO & C2D (48, 3) - C2D (48, 3) - C2D (48, 3) - C2D (48, 3) - V -  F (64) - F(75) & 44,304,571 \\
 
  NN-GF & F (256) - F (256) - F (256) - F (75) & 157,771\\
  \bottomrule
  \multicolumn{3}{l}{\parbox[t]{\textwidth}{\textbf{Layers:} [E] embedding ($\langle$embedding vector length$\rangle$); [C1D/C2D] convolution ($\langle$filter size, stride$\rangle$); [MP] 1-D max-pooling ($\langle$size$\rangle$) ; [AP] average pooling; [F] fully connected ($\langle$\# units$\rangle$); [D] dropout ($\langle$dropout value$\rangle$); [V] flatten.}}
\end{tabular}
}
\end{table*}  
 
In Fig.~\ref{fig:hyper_4k}, we show the impact of the hyper-parameters on the validation accuracy for 4,096-byte inputs (see supplementary materials for the 512-byte version). Each plot shows a violin plot with the distributions of validation accuracies (top), and the \emph{expected improvement} from the TPE. The best hyper-parameters are marked in dark green. Note that these values correspond to the best sampled network, and due to non-trivial interactions and stochastic nature of the process, may not necessarily correspond to seemingly best values suggested by TPE. The key observations are as follows:
 
\begin{enumerate}
  \item The convolution kernel size and stride seem to have the stringest and most consistent effect. Larger values lead to better performance regardless of the scenario.
 
  \item A single convolutional layer is typically not enough, two or three blocks are needed for best performance.
  
  \item In most scenarios, larger embedding sizes seem to be preferred, but good network with good performance can also be obtained for smaller embeddings - especially for smaller number of classes.
  
  \item For 4,096-byte blocks, stronger max-pooling seems to lead to better performance.
\end{enumerate}

Tab.~\ref{tab:models} shows the final architectures obtained for all scenarios for both 512 and 4096-byte inputs.

\section{Experimental Evaluation}
\label{sec:experiments}

\begin{table*}[t]
  \centering
  \caption{Detailed results for all 6 scenarios for 512 and 4096-byte blocks.}
  \resizebox{\textwidth}{!}{\begin{tabular}{C{0.75cm}C{1.5cm}C{3cm}C{25mm}C{6mm}C{6mm}C{6mm}C{6mm}C{7mm}C{7mm}C{7mm}C{7mm}C{1.2cm}}
 \toprule
 \textbf{Scen.} & \textbf{\#Classes} & \textbf{Classes} & \textbf{Composition$^1$} &  \multicolumn{2}{c}{\textbf{Accuracy}} & \multicolumn{2}{c}{\textbf{JPEG Acc.}} & \multicolumn{2}{l}{\textbf{Train. time [h]}} & \multicolumn{2}{l}{\textbf{Inf. time [$\frac{\text{ms}}{\text{block}}$]}} & \textbf{\#Samples} \\
 \textbf{} & \textbf{} & \textbf{} & \textbf{} & \textbf{512} & \textbf{4096} & \textbf{512} & \textbf{4096} & \textbf{512} & \textbf{4096} & \textbf{512} & \textbf{4096} &  \\
 \midrule
 \#1 & 75 & All & 75 & 65.6 & 77.5 & 83.5 & 86.3 & 7.51 & 12.53 & 0.043 & 0.146 & 7500k \\ \midrule
 \#2 & 11 & Bitmaps + RAW + Vector  + Video + Archives + Exe + Office + Published +  Human Readable + Audio + Other & (6) + (11) + (3) + (7) + (13) + (4) + (7) + (4) + (9)  + (7) + (4) & 78.9 & 89.8 & - & - & 1.98 & 7.55 & 0.036 & 0.294 & 1935k \\ \midrule
 \#3 & 25 & Bitmaps + RAW + Video + other & 6 + 11 + 7 + (51) & 87.9 & 94.6 & 93.3 &  98.9 & 4.26 & 10.16 & 0.063 & 0.227 & 2300k \\ \midrule
 \#4 & 5 & JPEG + RAW + Video + 5\_Bitmaps + Other & (1) + (11) + (7) + (5) + (51) &  90.2 & 94.1 & 98.6 & 99.1 & 0.50 & 4.95 & 0.043 & 0.400 & 1054k \\ \midrule
 \#5 & 2 & JPEG + non-jpegs & (1) + (74) & 99.0 & 99.2 & 99.3 & 99.2 & 0.32 & 1.89 &  0.045 & 0.134 & 1036k \\ \midrule
 \#6 & 2 & JPEG + non-jpegs & (1) + (11 + 3 + 2) & 99.3 & 99.6 & 99.5 & 99.7 & 0.41 & 4.34 & 0.047 & 0.259 & 1000k \\
 \bottomrule
  \multicolumn{13}{l}{\parbox[t]{2.0\columnwidth}{$^1$ Values in paranthesis represent the number of file types grouped into one class; values without paranthesis represent separate classes for each file type.}}\\
 \end{tabular}%
}
  \label{tab:scenarios}
\end{table*}

In this section, we present the results of our experimental evaluation. First, we introduce a novel dataset and distinguish several evaluation scenarios. Then, we describe 3 reference models, including a state-of-the-art tool \emph{Sceadan} and two alternative NN-based baselines. Finally, we report our results discuss the key limitations and causes of errors.

\subsection{Scenarios and Dataset}
\label{sec:scenarios}
\label{sec:data}

\begin{table}[t]
  \centering
  \caption{Computation time for global statistical \& bigram features.}
  \label{tab:feat}
  \resizebox{\columnwidth}{!}{\begin{tabular}{L{1.9cm}L{4cm}L{1cm}L{1cm}}
\toprule
\textbf{Feature} & \textbf{Description} & \multicolumn{2}{l}{\textbf{Runtime [ms/block]}} \\
& & \textbf{4096} & \textbf{512} \\ \midrule 
Kolgomorov Complexity & reduction of block length after bzip2/gzip compression \cite{BZ2,ZLIB} & 0.856 &  0.238 \\ \midrule 
Arithmetic Mean & self explanatory & 0.003 & 0.001 \\ \midrule 
Geometric Mean & self explanatory & 0.063 & 0.008 \\ \midrule 
Harmonic Mean & self explanatory & 0.048 & 0.007 \\ \midrule 
Standard Deviation & self explanatory & 0.046 & 0.006 \\ \midrule 
Mean absolute deviation & standard deviation using abs. value of difference instead of squares & 0.077 & 0.009 \\ \midrule 
Hamming Weight & average number of set bits in the block & 0.015 & 0.002 \\ \midrule 
Kurtosis of Byte Value & measure of peakedness in the byte value distribution graph & 0.110 &  0.015 \\ \midrule 
Skewness & Measure of asymmetry of the byte value distribution graph & 0.113 & 0.017 \\ \midrule 
Longest Byte streak & length of longest streak of repeating bytes in the block. & 0.981 &  0.156 \\ \midrule 
Low ASCII Range Freq & frequency of bytes in range: \texttt{0x00} - \texttt{0x1F} & 0.006 & 0.001 \\ \midrule 
Med ASCII Range Freq & frequency of bytes in range: \texttt{0x20} - \texttt{0x7F} & 0.013 & 0.002 \\ \midrule 
High ASCII Range Freq & frequency of bytes in range: \texttt{0x80} - \texttt{0xFF} & 0.018 & 0.002 \\ \midrule 
Shannon Entropy & $\Sigma \frac{1}{p}log_2\frac{1}{p}$ & 0.289 & 0.115 \\ \midrule
\multicolumn{2}{r}{\textbf{Total feature extraction time for NN-GF}} & \textbf{2.636} & \textbf{0.579} \\ \midrule
Bigram matrix & byte co-occurrence matrix (NN-CO) & \textbf{0.567} & \textbf{0.370} \\ 
\bottomrule
\end{tabular}}
\end{table}
 
Previous works relied mostly on small-scale private datasets or on \emph{GovDocs}~\cite{govdocs1} - an unbalanced corpus with 20 filetypes comprising 99.3\% and 43 filetypes comprising the remaining 0.7\% of the dataset. \emph{FiFTy} comes with a novel, large and balanced corpus, named FFT-75, comprising of 75 different filetypes with emphasis on multimedia types which dominate disk images  extracted from contemporary devices. 
 
We prepared the dataset by downloading public files from various public Internet sources and from our own collection. Instead of sharing the actual files, we sampled 102,400 small blocks for each filetype to obtain a balanced dataset well-suited for classification. Large and complex files cannot be recovered from the obtained blocks. For tiny text files potentially smaller than block size, we made sure to rely on open-source materials that explicitly permit sharing. To accommodate various file-system cluster sizes, we consider both 512 and 4,096-byte blocks. The blocks are shuffled to evenly distribute the filetypes within the training (80\%), validation (10\%) and hold-out testing (10\%) subsets. 

The included file-types are grouped by use-cases that include bitmaps (6 formats), RAW photographs (11), vector graphics (3), videos (7), archives (13), executable files (4), office documents (7), rich publications (4), human-readable text files (9), audios (7) and 4 others. More information, including detailed accuracy and misclassification rates are collected in Tab.~\ref{tab:misclassification}. To the best of our knowledge, the obtained corpus is the largest and most comprehensive to date. 
 
We distinguish \emph{6 scenarios} that correspond to various use cases and have different granularity (Tab.~\ref{tab:scenarios}). We focus on photo carving applications, where scenarios \#3 to \#6 are the most relevant:

\begin{itemize}
  \item \emph{\#1 (All; 75 classes)}: All filetypes are separate classes; this is the most generic case and can be aggregated into more specialized use-cases.
  
  \item \emph{\#2 (Use-specific; 11)}: Filetypes are grouped into 11 classes according to their use (see tags in 2$^{nd}$ column of Tab.~\ref{tab:misclassification}); this information may be useful for more-detailed, hierarchical classification or for determining the primary use of an unknown device.
  
  \item \emph{\#3 (Media Carver - Photos \& Videos; 25)}: Every filetype tagged as a bitmap (6), RAW photo (11) or video (7) is considered as a separate class;  all remaining types are grouped into one \emph{other} class. 
  
  \item \emph{\#4 (Coarse Photo Carver; 5)}: Separate classes for different photographic types: JPEG, 11 RAW images, 7 videos, 5 remaining bitmaps are grouped into one separate class per category; all remaining types are grouped into one \emph{other} class.  
  
  \item \emph{\#5 (Specialized JPEG Carver; 2)}: JPEG is a separate class and the remaining 74 filetypes are grouped into one \emph{other} class; scenario intended for analyzing disk images from generic devices.
  
  \item \emph{\#6 (Camera-Specialized JPEG Carver; 2)}: JPEG is a separate class and the remaining photographic/video types (11 RAW images, 3GP, MOV, MKV, TIFF and HEIC) are grouped into one \emph{other} class; scenario intended for analyzing SD cards from digital cameras.
\end{itemize}

Tab.~\ref{tab:scenarios} shows the class composition of each scenario, overall accuracy, JPEG accuracy, training and inference time along with number of samples in the particular dataset for all considered scenarios. The table shows that in general the accuracy improves as the number of classes decreases - except for scenario \#2 because it is more difficult than scenario \#3. Scenario \#2 groups filetypes by use, and not by internal structure, which may exhibit completely different statistics, e.g., the \emph{bitmap} class consists of TIFF, HEIC, BMP, GIF, PNG and JPEG which include images with no compression, as well as lossless or lossy compression. These distinct characteristics can be easily treated as separate classes in scenario \#3.

\subsection{Baseline Methods}
\label{sec:baseline}

To validate our approach, we compare \emph{FiFTy} against three baseline methods: the state-of-the-art tool \emph{Sceadan}, and two neural networks trained on global statistical features (\emph{NN-GF}) and co-occurrence features (\emph{NN-CO}). We test all methods on the largest scenario (\#1) for both 512 and 4,096-byte blocks. For \emph{Sceadan}, we used the open-source implementation and the provided training scripts~\cite{Sceadan}. For the remaining two baselines, we used the same TPE-based procedure (Section~\ref{sec:hyperparameters}) to optimize the models and their hyper-parameters. The final network architectures are collected at the bottom of Tab.~\ref{tab:models}.

\subsubsection{NN-CO} The co-occurrence or bigram features were commonly used for file-type classification~\cite{Wang_2018,Beebe_2013,Fitzgerald_2012,Karresand}. The feature is constructed from a matrix of byte co-occurrences, i.e., a matrix of shape (256, 256) with each element $a_{ij}$ corresponding to the frequency of occurrence of byte value $j$ after byte value $i$. Bigram features lead to good classification accuracy, but tends to be very sparse - especially for 512-byte blocks - and relatively slow to compute (Tab.~\ref{tab:feat}). As a result, they tend to be a performance bottleneck~\cite{Beebe_2013}. Due to the sparsity and excessive memory requirements, we down-sampled the matrices to (128, 128) by average pooling. The \emph{NN-CO} model comprises four 2-dimensional convolutional layers, and two dense layers operating on a flattened 1-dim representation. We keep the same kernel size and stride parameters for all layers, and treat them as hyper-parameters. Additional hyper-parameters include number of convolutional layers, presence of max pooling and the number of units in the hidden dense layer. 

\subsubsection{NN-GF} We use 14 global statistical features (Tab.~\ref{tab:feat}), commonly used in previous studies, as the input to the final baseline. Each feature maps an input block into a single scalar value. The features vary widely with respect to their computational effort, and also lead to performance bottlenecks in high-volume environments. The numbers reported in Tab.~\ref{tab:feat} correspond to feature extraction in Python 3.6 performed by calling low-level routines from popular external libraries (we relied on scipy 1.3.0 and numpy 1.16.4). This makes the runtime comparable to low-level C/C++ code. The classifier consists of four dense layers, and the number of units in all hidden layers was used as separate hyper-parameters. 

\begin{figure*}
  \centering
  \subfloat[][\emph{FiFTy} (65.6\%)]{\includegraphics[width=0.49\columnwidth]{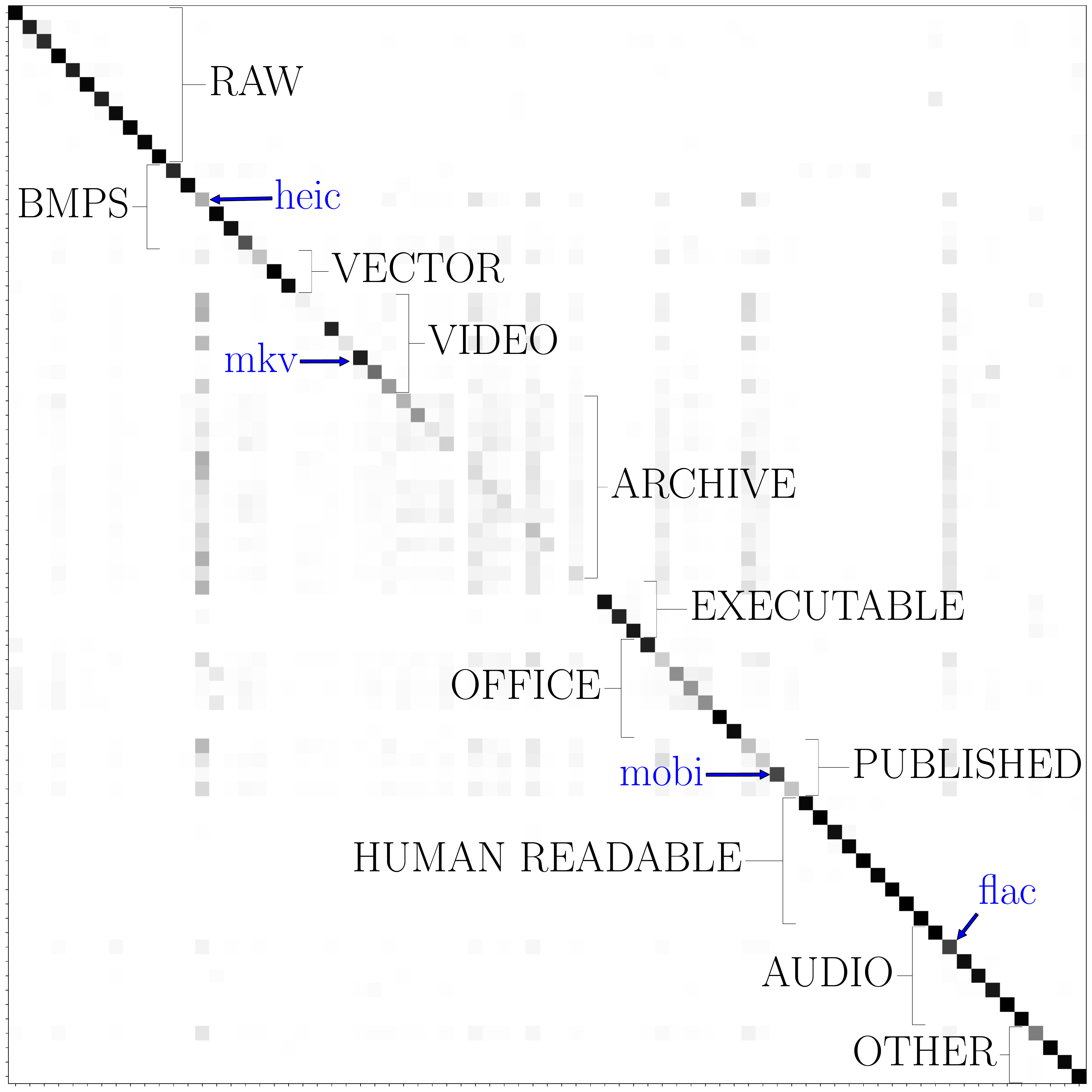}} \hspace{1pt}
  \subfloat[][Sceadan (57.3\%)]{\includegraphics[width=0.49\columnwidth]{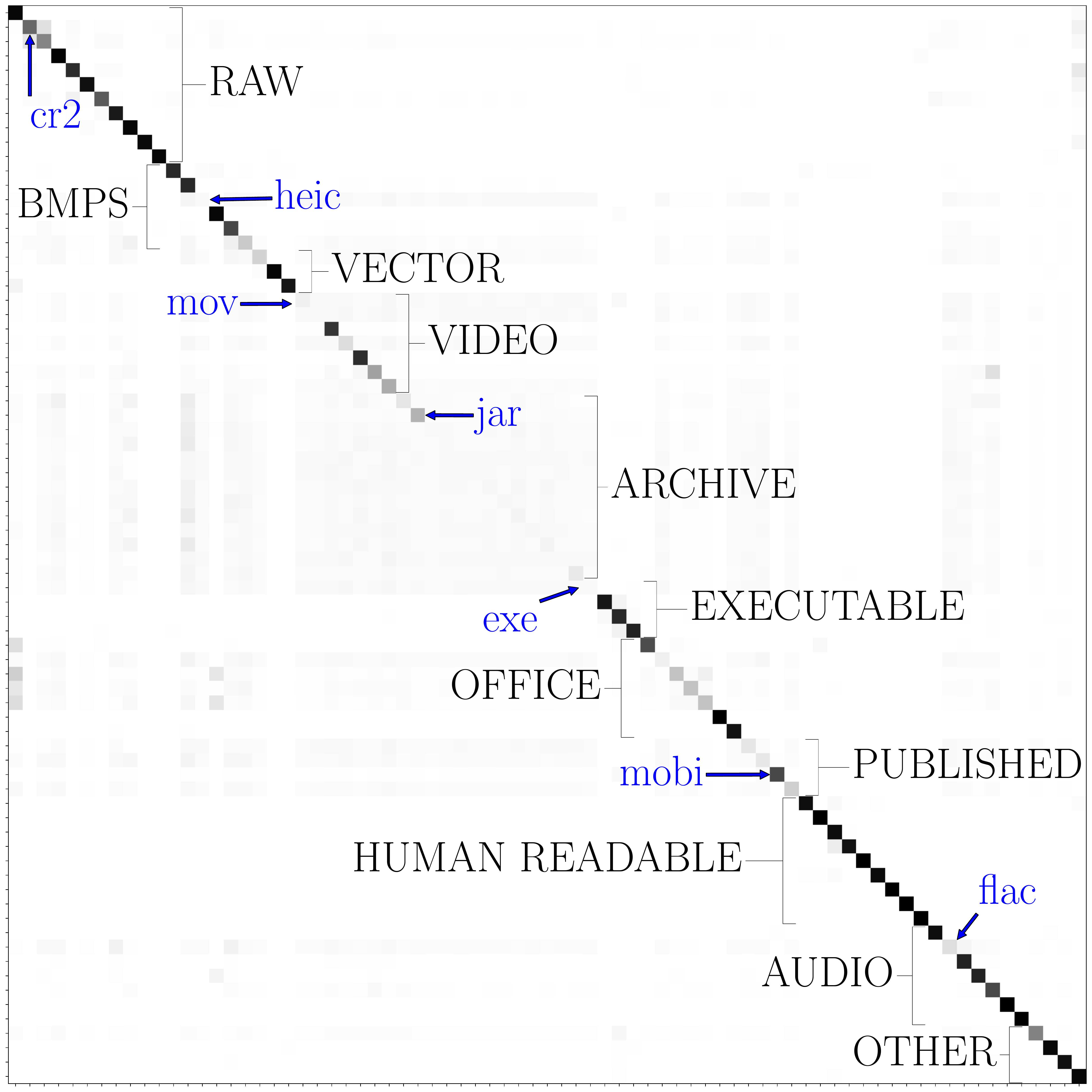}} \hspace{1pt}
  \subfloat[][NN-CO (64.4\%)]{\includegraphics[width=0.49\columnwidth]{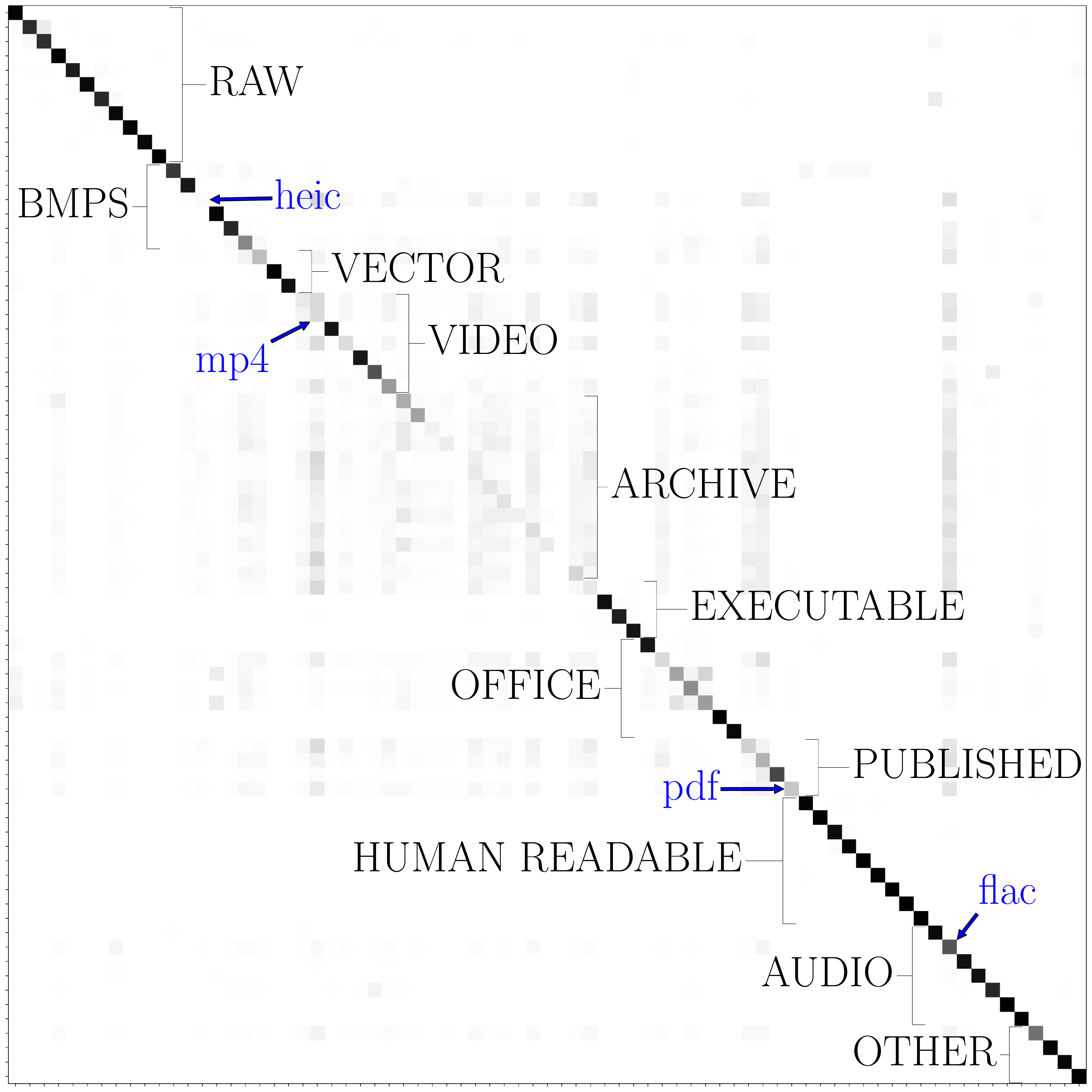}} \hspace{1pt}
  \subfloat[][NN-GF (45.4\%)]{\includegraphics[width=0.49\columnwidth]{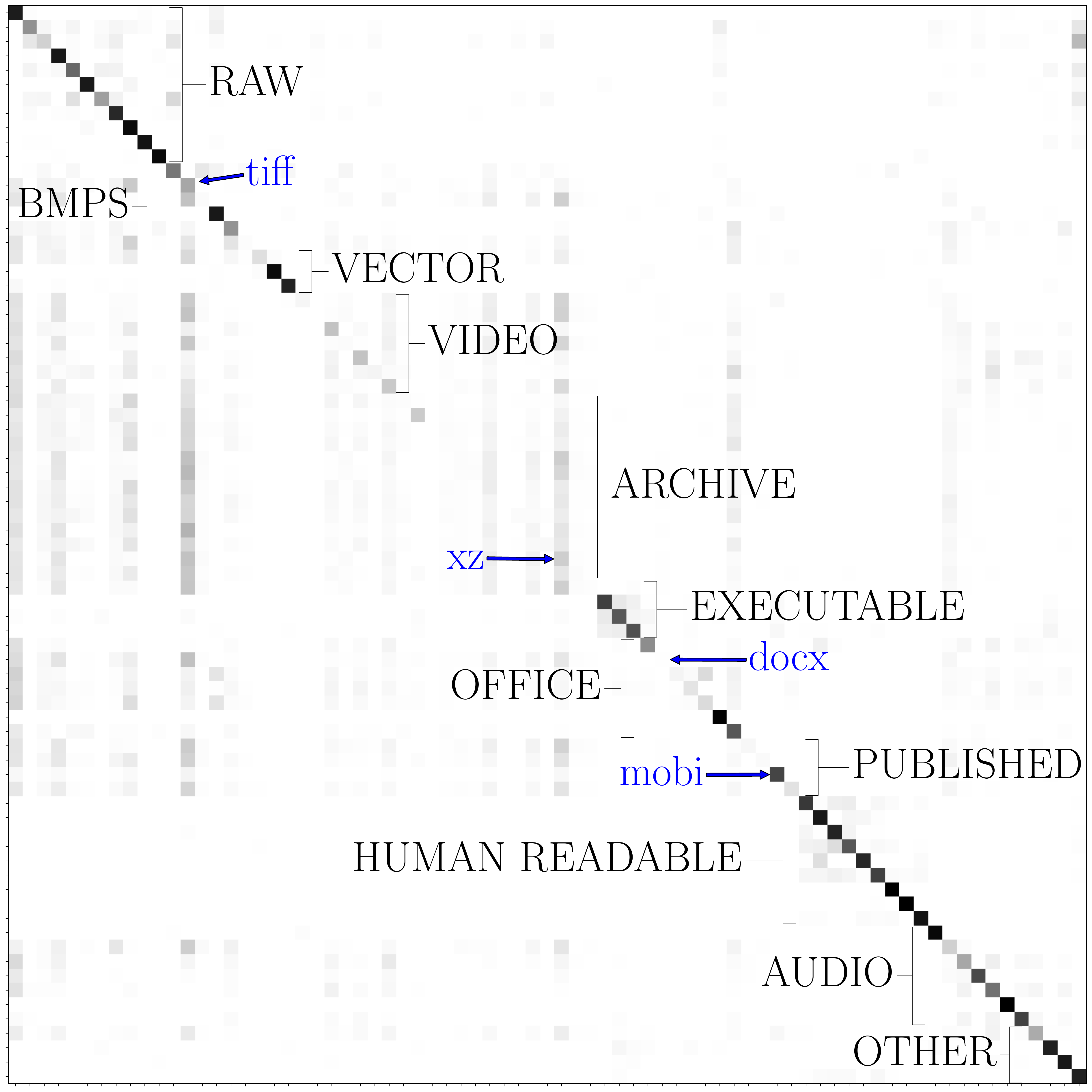}} 
  \\
  \subfloat[][\emph{FiFTy} (77.5\%)]{\includegraphics[width=0.49\columnwidth]{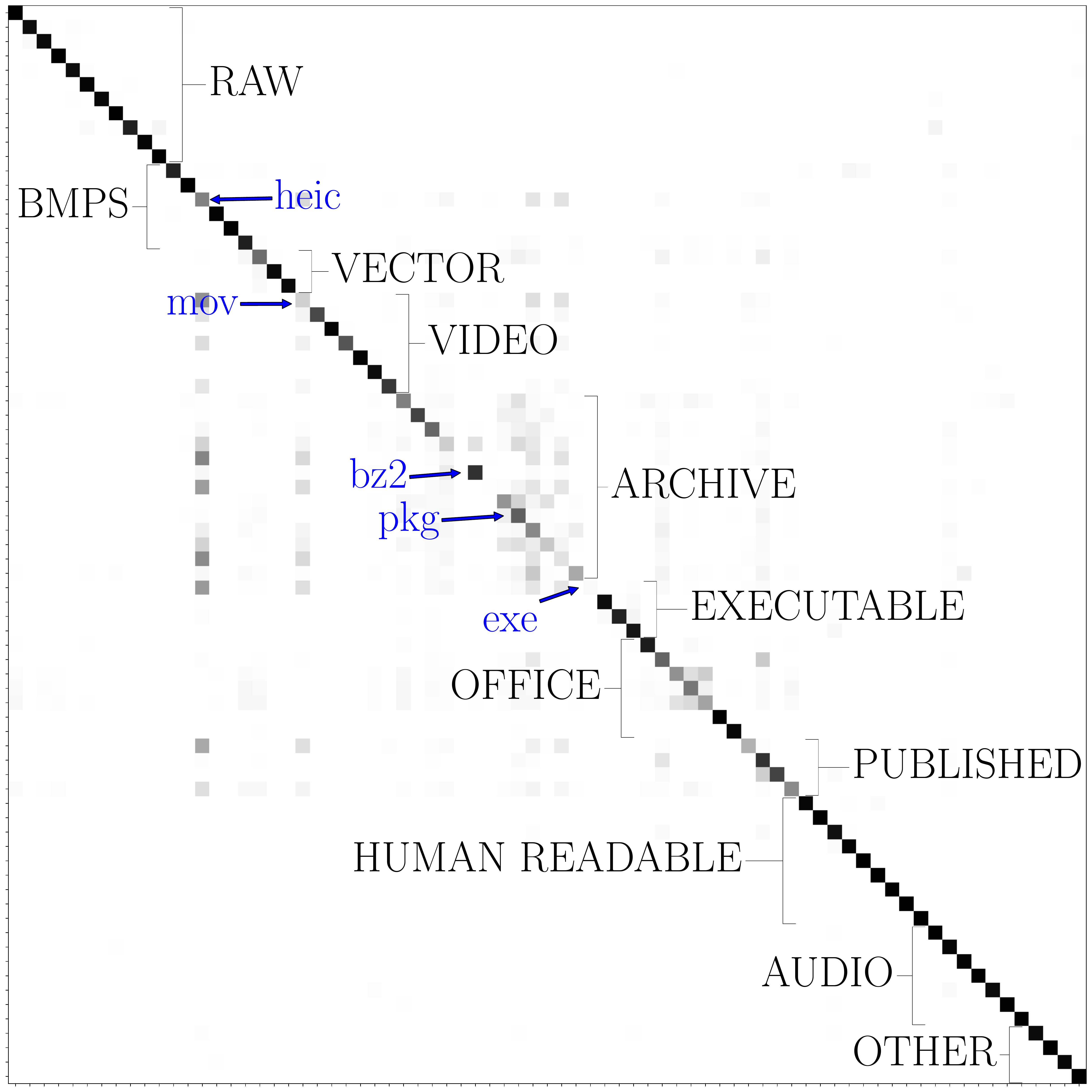}}  \hspace{1pt}
  \subfloat[][Sceadan (69.6\%)]{\includegraphics[width=0.49\columnwidth]{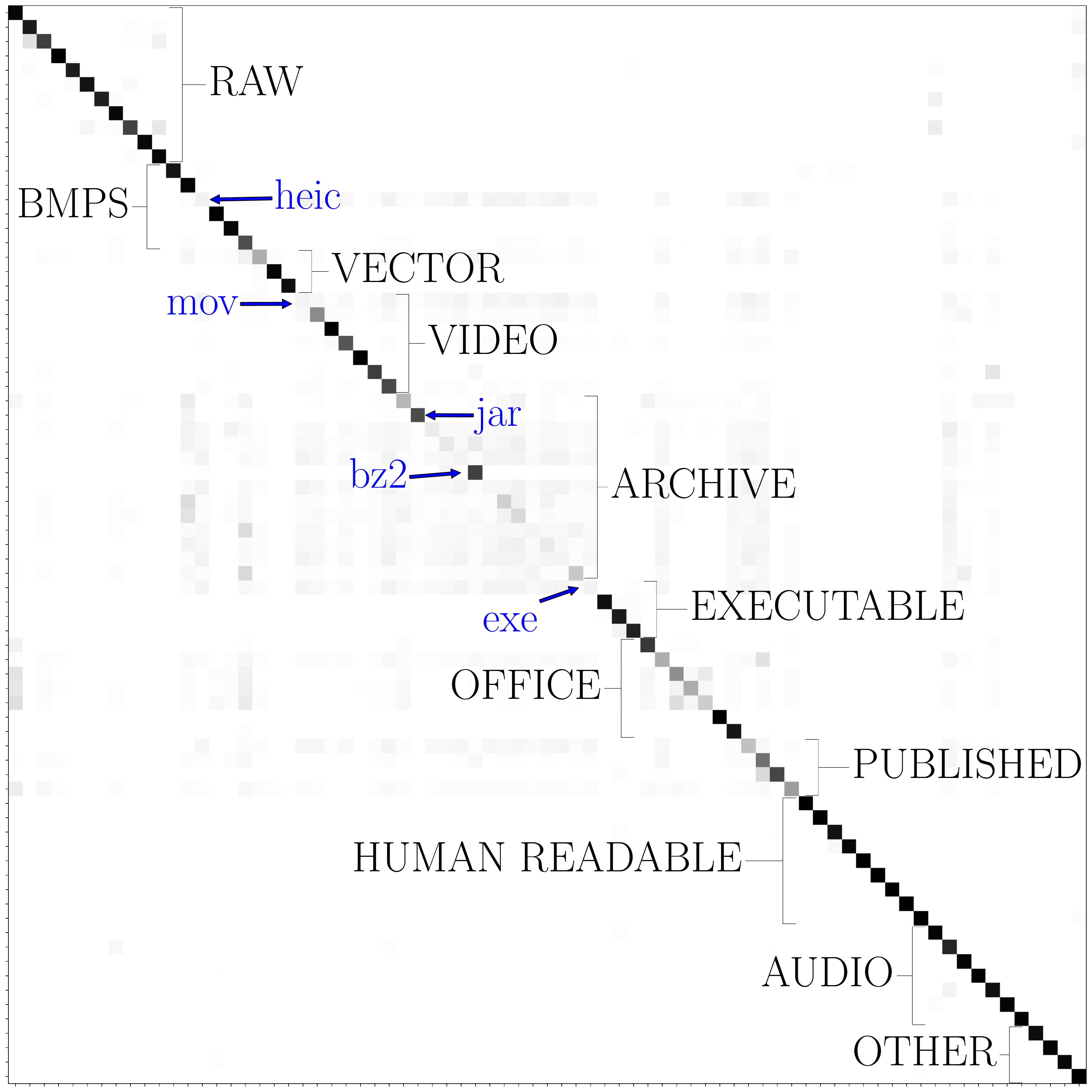}}  \hspace{1pt}
  \subfloat[][NN-CO (75.3\%)]{\includegraphics[width=0.49\columnwidth]{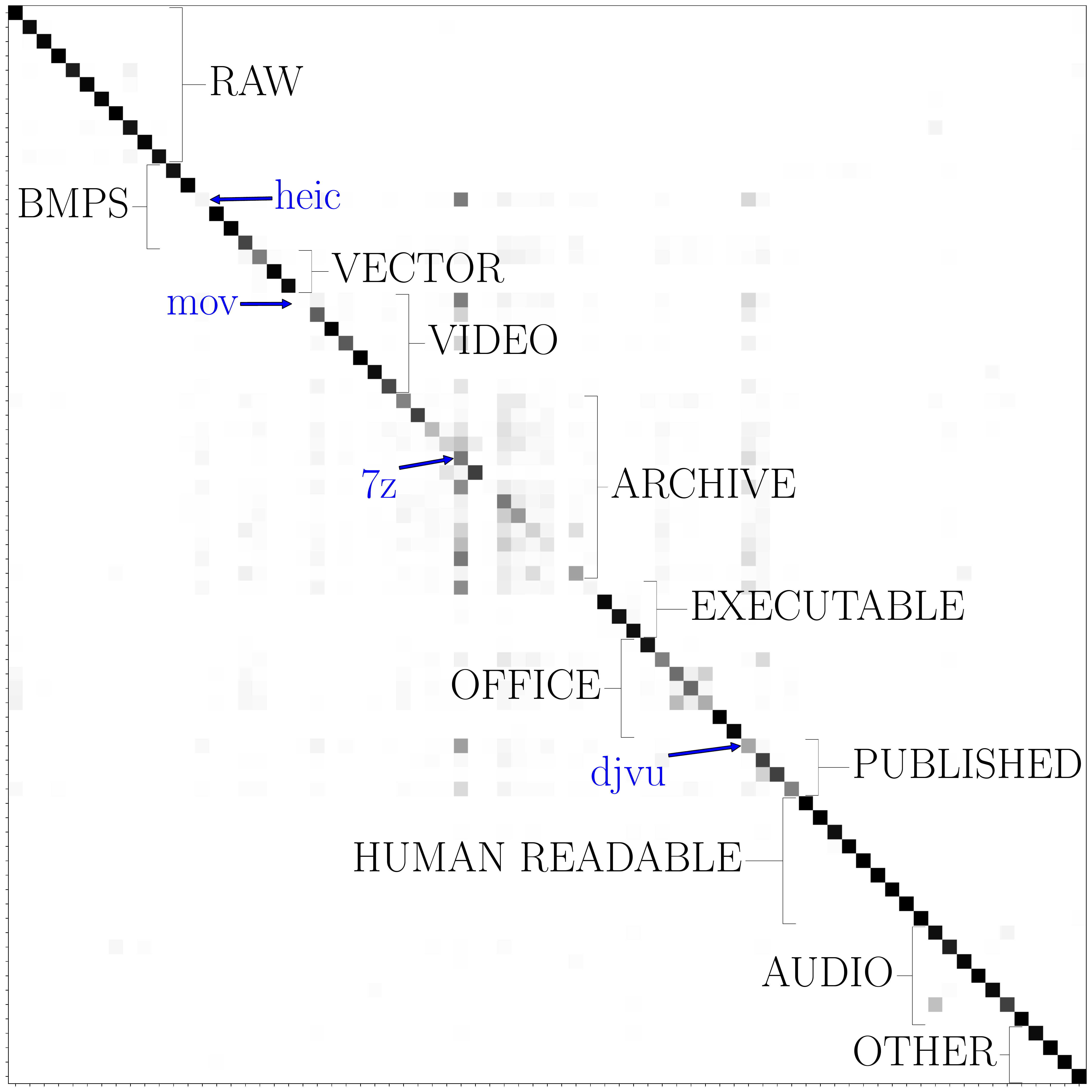}}  \hspace{1pt}
  \subfloat[][NN-GF (46.8\%)]{\includegraphics[width=0.49\columnwidth]{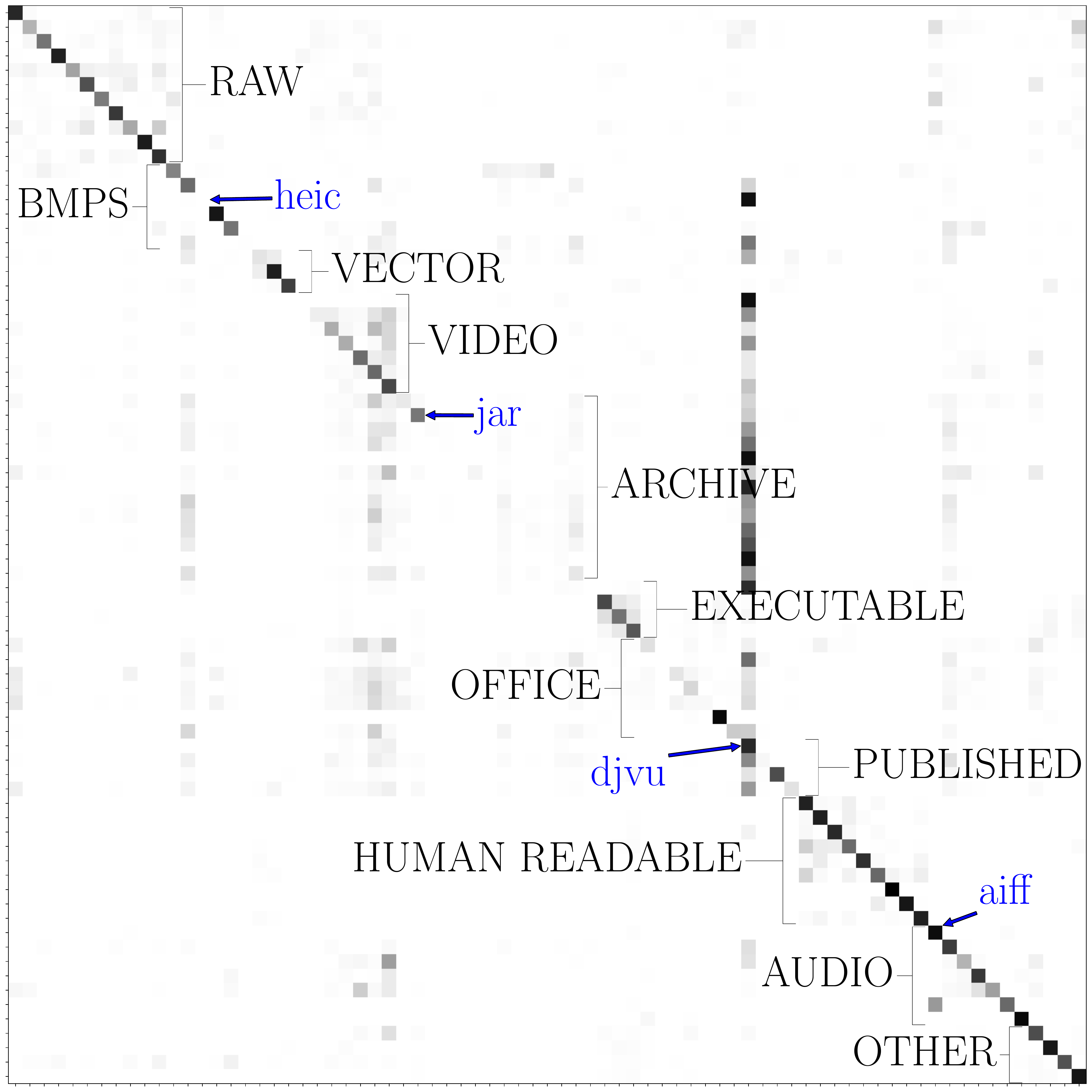}} 
  \caption{Heat maps of confusion matrices for scenario \#1 of \emph{FiFTy}, Sceadan~\cite{Beebe_2013}, NN-CO and NN-GF with block sizes of 512 bytes in subfigures \textbf{(a)-(d)} and 4096 bytes in subfigures \textbf{(e)-(h)}. Darker color means higher value.}
  \label{fig:comp}
\end{figure*}

\subsection{Evaluation Results}
\label{sec:eval_res}
Tab.~\ref{tab:prev} summarizes the overall accuracy and processing speed for \emph{FiFTy} and all of the baselines on scenario \#1 with 75 file-types. We report both the average accuracy and the JPEG identification accuracy. \emph{FiFTy} clearly outperforms all of the other methods - both with respect to accuracy and runtime. (Due to clearly inferior performance, we omit \emph{NN-GF} in the following discussion.) On 4,096-byte blocks, \emph{FiFTy} achieved average accuracy of 77.5\%, compared to 75.3\% and 69.0\% for the runner-ups (\emph{NN-CO} and \emph{Sceadan}) which were 7$\times$ and 14$\times$ slower, respectively. On the smaller 512-byte blocks, the accuracy deteriorates to 65.6\% vs. 64.4\% and 57.3\% with an even steeper speed-up gains - 20$\times$ and 34$\times$.

In terms of training time, \emph{FiFTy} was comparable to \emph{NN-CO} and significantly faster than \emph{Sceadan}, which took 44 hours for 512-byte inputs and over 10 days for 4096-byte inputs. Our model required $\approx$ 8 and 12 hours to train the final model on the entire dataset (on the slower P40 GPU), and between 24 - 36 hours for hyper-parameter search (depending on the scenario at hand). If needed, faster training times are possible - a sub-set of the training set and a slightly simpler model can cut down the training time to around 1-2 h with only minor degradation in accuracy (Fig.~\ref{fig:data-size-impact}).

Fig.~\ref{fig:comp} compares confusion matrices for all of the models. Due to a large number of classes, the included figure depicts groups of file-types and marks selected individual file-types of interest. For detailed file-type-level results, please refer to supplementary materials or Table~\ref{tab:misclassification} which summarizes the most common mistakes. \emph{FiFTy} delivered the best accuracy for 43 out of 75 individual file-types. The \emph{NN-CO} was the winner for 27 and \emph{Sceadan} for 4 (for 4,096 blocks). Whenever \emph{FiFTy} lost, the difference was marginal.

\subsubsection{Ensemble of \emph{FiFTy} and \emph{NN} baselines}

 We consider ensemble learning to assess whether inclusion of hand-engineered features could improve the accuracy. By averaging the probabilities prodicted by \emph{FiFTy}, \emph{NN-CO}, and \emph{NN-GF} we can achieve 78.7\% on 4,096-byte inputs - a marginal improvement not worth the enormous computational effort. We also tried training a simple combination NN, but did not observe any further improvements.

\subsubsection{Additional Scenarios}

Tab.~\ref{tab:scenarios} summarizes \emph{FiFTy} evaluation results for all test scenarios. As the number of classes decreases, we obtain increasingly more reliable results, reaching 94.6
\% accuracy for 25 multimedia file-types and 99.7\% accuracy for the most constrained JPEG-vs-others scenario. To the best of our knowledge, these are the best such results to date. The best previously reported JPEG identification accuracy was 99.2\% on a dataset with only 4 very different file-types~\cite{Karresand}. Our classifier considered 16 other photographic and video types in the alternative class, which makes it suitable for photo carving applications. Even for the detailed multimedia scenario \#3 (25 file-types), we achieved JPEG classification accuracy of 98.9\%.

\begin{table}[t]
  \centering
  \caption{Inference Time (ms/block) of \emph{FiFTy} for various scenarios tested on a Tesla V100, P40, K80 GPU and 4 E5-2690 CPUs.}
  \label{tab:gpu-comparison}
  \resizebox{\columnwidth}{!}{%
\begin{tabular}{C{1.2cm}C{0.65cm}C{0.65cm}C{0.65cm}C{0.65cm}C{0.65cm}C{0.65cm}C{0.65cm}C{0.875cm}}
\toprule
\textbf{Scenario} & \multicolumn{4}{c}{\textbf{512-byte blocks}} & \multicolumn{4}{c}{\textbf{4,096-byte blocks}} \\ \midrule 
\textbf{GPUs} & \textbf{V100} & \textbf{P40} & \textbf{K80} & \textbf{CPU} & \textbf{V100} & \textbf{P40} & \textbf{K80} & \textbf{CPU} \\ \midrule
\#1 & 0.043 & 0.058 & 0.181 & 3.102 & 0.146 & 0.256 & 0.947 & 16.836 \\ 
\#2 & 0.036 & 0.047 & 0.126 & 1.907 & 0.163 & 0.294 & 1.014 & 20.519 \\ 
\#3 & 0.063 & 0.086 & 0.353 & 5.694 & 0.119 & 0.227 & 0.628 & 11.526 \\ 
\#4 & 0.043 & 0.057 & 0.191 & 2.859 & 0.245 & 0.400 & 1.705 & 33.900 \\ 
\#5 & 0.045 & 0.057 & 0.232 & 3.893 & 0.097 & 0.134 & 0.541 & 12.569 \\ 
\#6 & 0.047 & 0.059 & 0.105 & 1.330 & 0.137 & 0.259 & 0.901 & 17.519 \\
\bottomrule
\end{tabular}%
}
\end{table} 

\subsubsection{Runtime Evaluation}

The classifier comprising the heart of \emph{FiFTy} is a neural network implemented in a high-level framework Keras. The model can take advantage of GPUs to speed up computations. In this experiment, we assess the inference speed on several popular server GPUs (nVidia Tesla V100, P40 and K80). The results are collected in Tab.~\ref{tab:gpu-comparison}. Even with parallel processing, general-purpose CPUs are not competitive for high-volume applications. Adoption of GPUs significantly speeds-up the computations. On the fastest considered GPU and with 4,096-byte blocks, we achieved the speed of 214 Mbps for the largest scenario with 75 classes, and up to 322 Mbps for the simplest 2-class problem.

\subsection{Misclassification Analysis}

The last two columns in Tab.~\ref{tab:misclassification} show a misclassification breakdown for all of the considered file-types. For brevity, we show only top-2 types most confused with each given file-type (for scenario \#1/4096). There are several reasons for such errors. We list the most interesting observations below:

\begin{itemize}
  \item \emph{HEIC \& MOV:} HEIF is a new image format used in the recent iPhones~\cite{HEIF}. It is essentially a container for still photos encoded using the intra-frame mode of the HEVC video codec~\cite{sullivan2012overview}. This leads to misclassification errors, since HEIC photographs are essentially small movies and tend to be confused with actual MOV videos. Moreover, due to high entropy, both formats are sometimes confused with archives - especially the XZ and RAR formats. 
    
  \item \emph{Archives:} Data compression (whether lossy or lossless) reduces redundancy and stores the same information using fewer bits. Archive file-types are characterized by high entropies, thus minimizing statistical artifacts which could be exploited to identify them. Almost all included types suffer from this problem. Although the proposed \emph{FiFTy} improves upon previous methods, the accuracy still averages around only 39\%. The only format with reasonable accuracy was BZ2, which obtained 80\%.

  \item \emph{Compound filetypes:} Many filetypes, e.g., PDF, PPT, KEY, DOC, MOBI are containers and may have embedded objects of other types, e.g., JPEG or other bitmaps, audio, video, etc. This confuses all classifiers, but seems to be mostly a data labeling problem. In practice, such segments should probably be carved using dedicated tools for the embedded objects. However, the container file should most likely be carved as well. More complex logic will be needed both for carving such data as well as properly labeling the training data for file-type classifiers.
  
  \item \emph{Presentation File-Types:} We observed a cluster of mutual misclassifications for common presentation file-types (PPT, PPTX and KEY). Interestingly, we didn't observe simialar relationships between document and spreadsheet types. To some extend the problem may stem from file-type conversion (see below).

  \item \emph{Errors stemming from type conversion:} Some files in our dataset were generated by converting from other files of a similar type (to reach the required number of samples). We observed elevated mutual misclassification rates for some of the pairs (WAV \& AIFF, KEY \& PPTX, PPT \& PPTX and MOBI \& EPUB). Other pairs (RTF \& TXT, HTML \& MD or TEX \& MD) remain unaffected. More information about the conversion can be found in the documentation of the dataset.
\end{itemize}

\subsection{Generalization}

We tested generalization capabilities of \emph{FiFTy} on the GovDocs~\cite{govdocs1} corpus and a file-system memory dump from a DFRWS\footnote{http://old.dfrws.org/2006} carving challenge.

\begin{figure}
  \centering
  \subfloat[][GovDocs (\#1/4,096 - 55.9\%)]{\includegraphics[width=0.49\columnwidth]{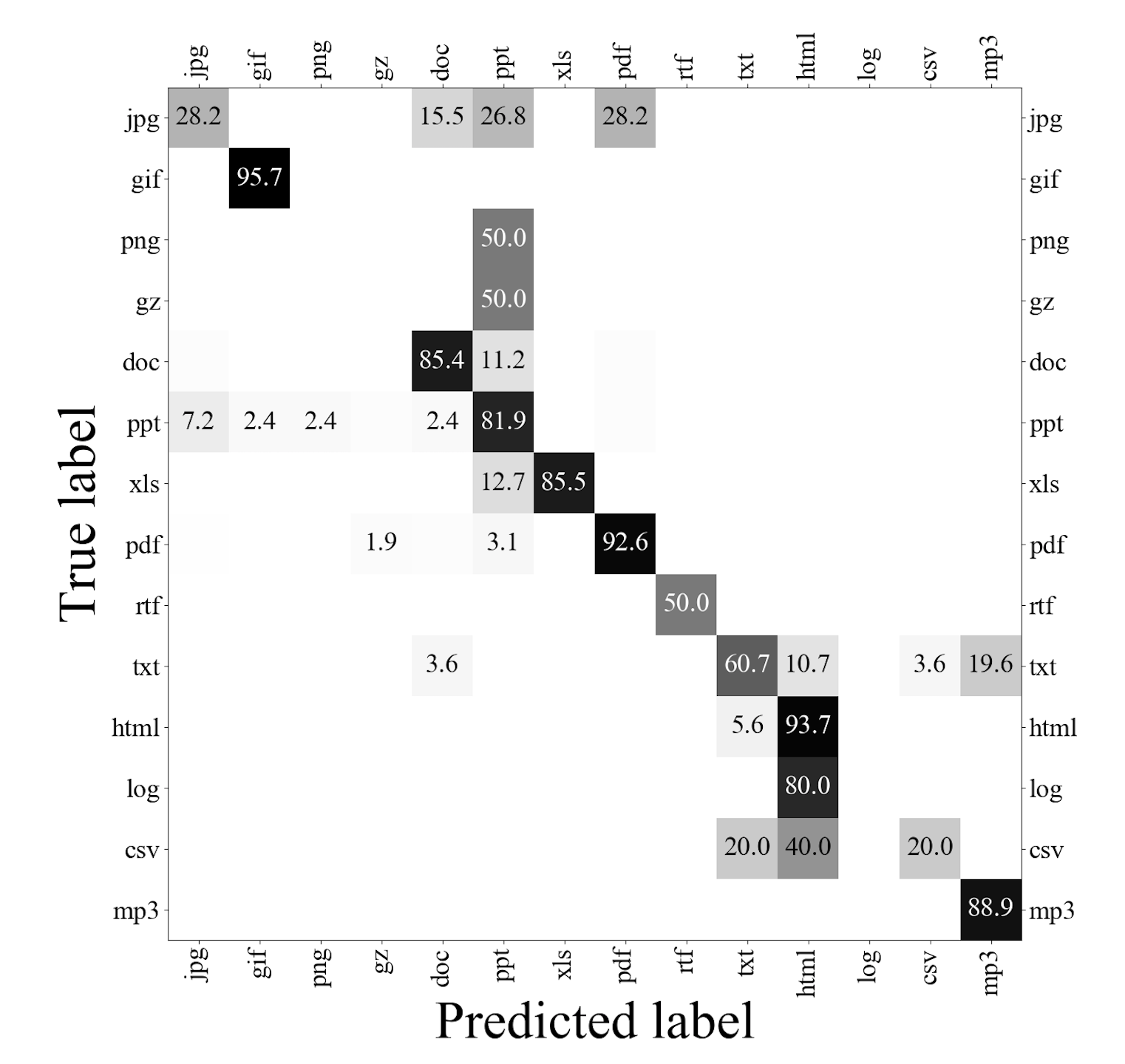}} \hspace{1pt}
  \subfloat[][GovDocs (\#4/4,096 - 88.0\%)]{\includegraphics[width=0.49\columnwidth]{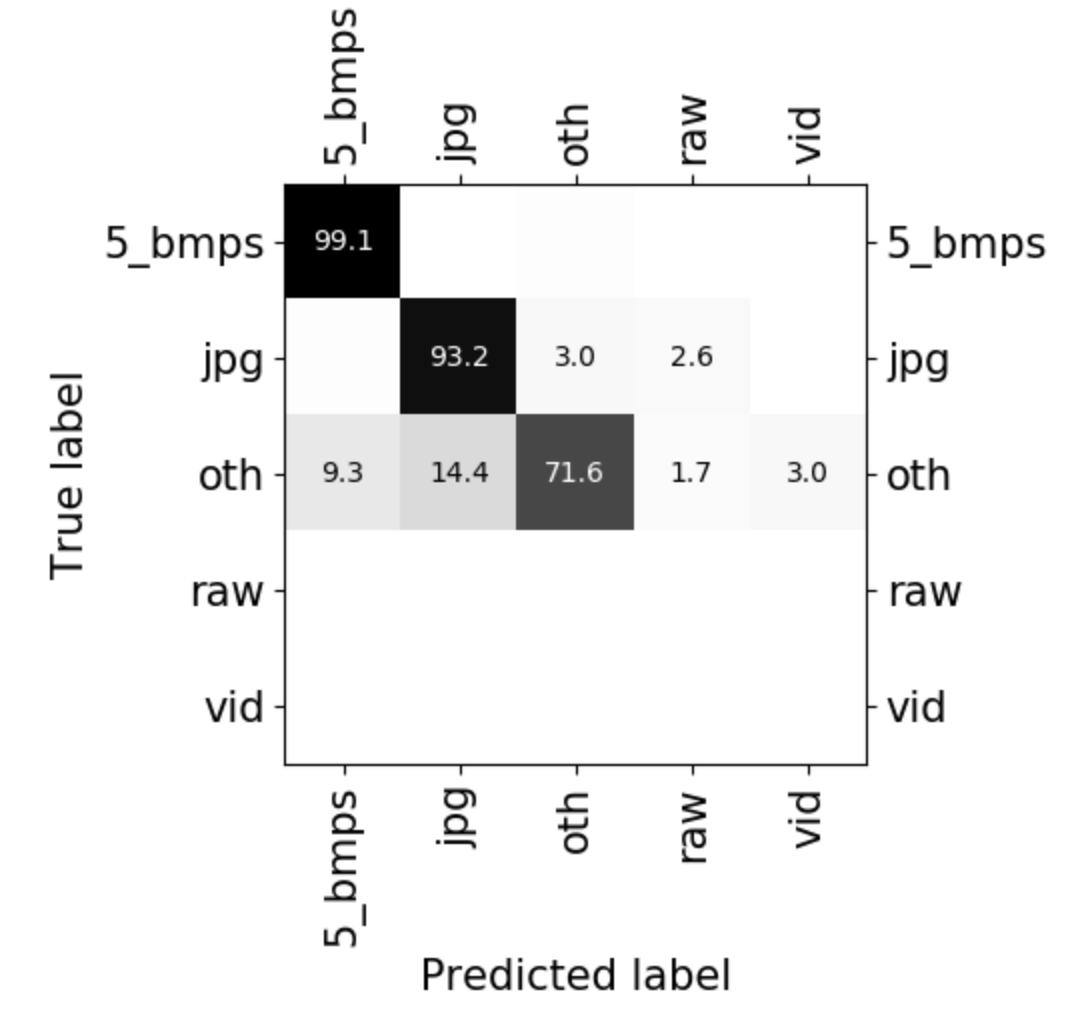}} \hspace{1pt} \\ 
  \subfloat[][DFRWS (\#5/512 - 86.8\%)]{\includegraphics[width=0.49\columnwidth]{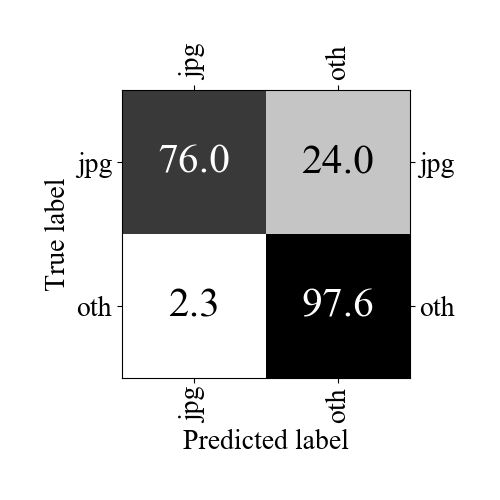}} \hspace{1pt}
  \subfloat[][DFRWS (\#6/512 - 78.1\%)]{\includegraphics[width=0.49\columnwidth]{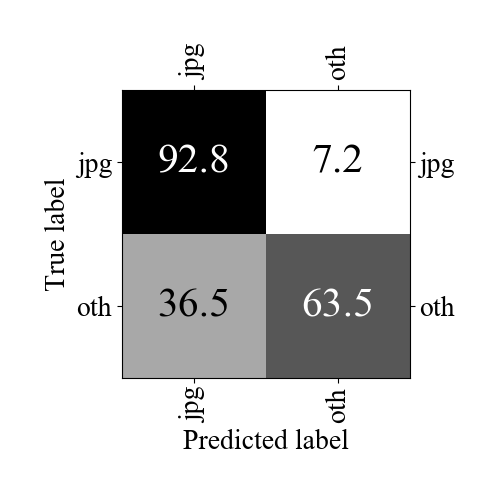}} 
  \caption{Confusion matrices: on Govdocs~\cite{govdocs1} for scenario \#1 and \#4 (ab) with block size of 4096 bytes; and on the DFRWS image for scenario \#5 and \#6 (cd) with block size of 512 bytes. Darker color means higher value.}
  \label{fig:generalization}
\end{figure}

\paragraph{GovDocs} GovDocs~\cite{govdocs1} is a large corpus of files submitted to US government websites. We took a portion of that corpus with sufficient number of testing examples and overlap with our file-types: CSV, GIF, GZ, HTML, JPG, LOG, XLS, PPT, DOC, RTF, PDF, MP3, PNG and TXT. We used 4,096-byte blocks for preparing a shuffled disk image. Table~\ref{tab:general} summarizes the validation accuracy for all scenarios, and Fig.~\ref{fig:generalization}ab show the confusion tables for two example scenarios. 

Please note that we report accuracy on random memory blocks derived from the corpus and not on entire files as is sometimes done by other studies. Moreover, the displayed rows may not add up to 100\% because we only show the common classes between GovDocs and scenario \#1. For example, none of DOCX, PPTX and XLSX filetypes are present in GovDocs. Therefore, if a PNG block is predicted to be an APK block, then it is missing in the presented confusion matrix. Similarly for scenario \#4, there are no raw images or video files, therefore their rows are empty. Absent classes can potentially indicate classification errors.

The key observations about the mistakes made by the classifier are as follows:
  \begin{itemize}
    \item In scenario \#1, JPG is confused with PDF, PPT and DOC because most of these files had JPG images embedded in them. Similar problems occur for PNG.
    \item In scenario \#1, RTF and TXT files are confused with other file-types because the samples in Govdocs were very often structured and mislabeled (e.g., tabular data, or postscript).
    \item In scenario \#1, CSV could be confused with HTML because the CSV samples have more quotation marks (\texttt{0x22}) than commas (\texttt{0x2c}), which is similar to HTML files.
  \end{itemize}

\paragraph{DFRWS} We took a disk image from the DFRWS-2006 challenge. The file system has cluster size of 512 bytes and contained XLS, GZ, JPG, HTML, DOC and TXT files deliberately shuffled to make carving harder. Table~\ref{tab:general} summarizes the validation accuracy for all scenarios, and Fig~\ref{fig:generalization}cd show the results on two example scenarios.

A key observation for the above two samples is that although the classes of both scenarios \#5 and \#6 are the same, we see different accuracies because the \emph{other} class consists of different sets of file-types (refer Sec.~\ref{sec:scenarios}). It is important to note that the classifier for scenario \#6 was not trained on any other file-types in DFRWS image, except JPG. Scenario \#5 - other class consists of all 74 file-types and scenario \#6 - other class consists of 16 multimedia file-types. Classifier for scenario \#5 learned better what is not JPG and for \#6 learned better what is JPG.

Table~\ref{tab:general} presents a summary of results on the above two samples for all the six scenarios. We observe that the tool gives all results comparable results to the original testing accuracy on FFT-75 for scenarios \#3 - \#6. All these scenarios are related to multimedia file-types. Hence, the tool does show evidence that it is efficient and accurate asset for multimedia carving and related applications.

\begin{table}[t]
  \centering
  \caption{Accuracy on GovDocs and DFRWS for all scenarios }
  \label{tab:general}
  \resizebox{\columnwidth}{!}{\begin{tabular}{lrrrrrr}
    \toprule
    \textbf{Data\textbackslash  Scenario} & \textbf{\#1} & \textbf{\#2} & \textbf{\#3} & \textbf{\#4} & \textbf{\#5} & \textbf{\#6} \\
    \midrule 
    \textbf{GovDocs (4096)} & 55.9\% & 45.2\% & 83.9\% & 88.0\% & 88.3\% & 82.3\% \\
    \textbf{DFRWS (512)}  & 41.4\% & 46.5\% & 60.8\% & 71.9\% & 86.8\% & 78.1\% \\ 
    \bottomrule
\end{tabular}

}
\end{table}

\subsection{Limitations and Future Work}

The presented neural network architecture reflects both the effects of hands-on experimentation and automatic hyper-parameter tuning. Out of necessity, we explored only a small fraction of the design space. It is certainly possible to expect that better architectures could be found using modern architecture search~\cite{liu2018progressive} or even random sampling methods~\cite{xie2019exploring}. These methods still remain an area of active research, and its application for the problem at hand is left for future work. 

The FFT-75 dataset used for training suffers from some dependency problems. There are a lot of DOC, PDF and PPT files that contain images embedded in them. This makes it difficult for the classifier to learn and generalize. An approach to solving this could be to carve the images and other embeddable data-types out of these files and train on the dataset prepared from these files.

Our current hyper-parameter search was driven mainly by validation accuracy. This sometimes leads to surprisingly high model complexity for seemingly simpler problems (Tab.~\ref{tab:scenarios}). In the future work, we're planning to investigate the relationship between model complexity (e.g., parameter count), computational effort and accuracy. The problem requires a separate treatment since it heavily depends on the target runtime environment, and both hardware platforms and tensor libraries can result in widely different behavior~\cite{shi2016benchmarking}.

Additional improvements in classification accuracy can be expected by adopting more complex logic to deal with embedded filetypes as well as with the dependencies between labeling decisions among neighboring blocks. For example, random field models could be used to reach an agreement regarding a larger chunk of data based on weak predictions of individual blocks~\cite{lafferty2001conditional}. This information could further be combined with the locations of known headers/footers. 

Finally, we emphasize that in real applications additional care may be needed for certain filetypes to accommodate the expected content characteristics. For example, statistical properties of video files may depend not only on the container, but also on the audio/video codecs. In our study, we relied on material that we captured using common camera-enabled devices (e.g., GoPro, dash cameras). We used default codec settings which should be most applicable in common media carving applications. Our dataset included mainly MPEG4, H.264, H.265, VP8, VP9 and OGV codecs with different distribution across containers. We leave a more detailed study of this phenomenon for future work.

\section {Conclusions}
\label{sec:conclusions}

Our study addressed the problem of large-scale file fragment type identification. We explored neural network models, which dispense with explicit feature engineering - a bottleneck from both development and computation perspectives. Our architecture is based on 1-D convolutions. We accept raw blocks of bytes as input, and embed them in a trainable real-valued latent space. By changing a few hyper-parameters, our architecture can be easily adapted to various applications. We used the Tree-structured Parzen Estimator~\cite{bergstra2011algorithms} to analyze the impact of hyper-parameters and find their optimal values. The resulting models are included in \emph{FiFTy}, an open source tool for file-type identification. 

Our evaluation is based on a novel dataset with 75 popular file-types (available publicly at~\cite{kfxw-8084-19}). We focused on multimedia and photographic types which commonly occur in data carving applications. Based on the obtained results, we can conclude that neural networks are an extremely effective and flexible tool for file-type identification. The trained classifiers consistently outperformed all baseline methods. Moreover, thanks to modern software frameworks and hardware accelerators, classification can run an order of magnitude faster.
 
\begin{table*}[ht!]
  \centering
  \caption{File-type breakdown of classification accuracy for all models; misclassification reported for \emph{FiFTy} on 4,096-bytes inputs.}
  \label{tab:misclassification}
  \resizebox{0.93\textwidth}{!}{%
  \begin{filecontents*}{data/misclassification.csv}
Filetype,Tag,Co_occur_512,Feature_512,Sceadan_512,CNN_512,Co_occur_4k,Feature_4k,Sceadan_4k,CNN_4k,top1,top2
ARW,Raw,97.8,88.6,97.3,\textbf{97.8},98.8,84.7,97.5,\textbf{98.9},JPG (0.4),MOV (0.1)
CR2,Raw,82.4,43.1,58.2,\textbf{86.7},\textbf{95.3},30.0,88.9,94.8,NEF (1.1),DNG (0.6)
DNG,Raw,81.6,18.1,47.7,\textbf{82.7},\textbf{96.7},54.6,75.3,96.1,CR2 (2.0),NEF (0.4)
GPR,Raw,98.7,89.8,98.1,\textbf{99.2},99.5,87.2,99.3,\textbf{99.9},CR2 ($<$0.1),PEF ($<$0.1)
NEF,Raw,\textbf{89.3},59.6,82.5,87.7,89.0,36.1,88.3,\textbf{95.3},NRW (0.9),JPG (0.9)
NRW,Raw,96.1,89.4,93.9,\textbf{96.9},96.7,68.7,92.2,\textbf{97.7},RAF (0.9),3FR (0.3)
ORF,Raw,84.0,37.7,63.7,\textbf{86.3},\textbf{97.6},51.9,87.3,96.3,FLAC (1.4),NEF (0.7)
PEF,Raw,\textbf{96.1},84.3,88.8,95.1,98.9,77.5,96.8,\textbf{99.1},NEF ($<$0.1),ORF ($<$0.1)
RAF,Raw,97.9,95.4,96.6,\textbf{98.3},\textbf{91.3},34.0,73.6,87.1,FLAC (4.6),3FR (4.1)
RW2,Raw,\textbf{96.6},92.4,95.5,96.5,97.9,89.3,95.9,\textbf{97.9},JPG (1.0),PPT (0.2)
3FR,Raw,99.4,96.0,96.6,\textbf{99.6},92.6,81.0,97.2,\textbf{99.5},DNG (0.2),TIFF (0.2)
JPG,Bitmap,77.5,53.2,83.8,\textbf{83.5},\textbf{92.5},48.6,91.8,86.3,PPT (3.9),PPTX (2.3)
TIFF,Bitmap,90.0,34.5,83.8,\textbf{96.1},98.6,57.9,97.7,\textbf{99.0},APK (0.2),PPT (0.2)
HEIC,Bitmap,2.0,1.8,2.9,\textbf{31.7},4.0,0.0,5.2,\textbf{49.5},MOV (15.3),XZ (11.6)
BMP,Bitmap,97.5,90.1,96.1,\textbf{98.0},\textbf{99.4},91.3,98.7,98.4,3FR (0.5),DLL (0.3)
GIF,Bitmap,84.5,42.3,72.3,\textbf{93.5},98.5,54.3,96.2,\textbf{99.1},MSI ($<$0.1),OGV ($<$0.1)
PNG,Bitmap,47.6,1.5,21.7,\textbf{67.2},72.9,0.4,69.5,\textbf{88.0},PKG (1.7),RAR (1.2)
AI,Vector,\textbf{25.5},12.6,17.4,23.3,50.9,10.6,31.7,\textbf{57.7},EPUB (6.9),PKG (5.8)
EPS,Vector,\textbf{98.8},95.2,97.8,98.7,\textbf{98.0},88.2,97.4,95.8,AI (3.4),PDF (0.4)
PSD,Vector,93.7,86.2,91.2,\textbf{95.3},95.6,75.3,94.4,\textbf{95.9},AI (2.1),JPG (0.4)
MOV,Video,\textbf{8.7},3.6,6.1,6.1,0.0,0.0,4.8,\textbf{18.5},HEIC (43.1),RAR (13.0)
MP4,Video,\textbf{14.8},0.0,2.8,1.6,62.2,7.0,45.5,\textbf{71.8},HEIC (11.2),XZ (2.9)
3GP,Video,\textbf{91.6},23.1,79.3,85.6,\textbf{98.8},32.1,98.7,98.1,AVI (0.6),DMG (0.2)
AVI,Video,13.8,1.7,\textbf{14.7},10.9,64.9,32.6,66.4,\textbf{67.1},HEIC (13.5),MOV (6.2)
MKV,Video,\textbf{90.5},23.5,82.3,88.0,\textbf{99.1},57.5,98.9,98.4,OGV (0.4),DMG (0.2)
OGV,Video,\textbf{67.4},4.6,37.3,57.3,94.5,59.0,76.0,\textbf{94.9},OGG (1.3),MSI (0.8)
WEBM,Video,39.6,21.4,33.3,\textbf{39.8},71.5,71.6,71.2,\textbf{78.1},HEIC (10.0),MOV (3.2)
APK,Archive,\textbf{32.5},1.5,10.0,29.9,48.3,9.0,29.0,\textbf{49.7},PKG (11.5),PPT (3.5)
JAR,Archive,36.3,20.4,28.9,\textbf{41.2},\textbf{74.8},53.4,70.2,73.8,GZ (6.1),RPM (4.1)
MSI,Archive,5.1,0.3,3.3,\textbf{10.1},26.6,1.5,7.6,\textbf{60.1},PKG (5.2),EPUB (3.0)
DMG,Archive,7.0,1.1,3.4,\textbf{18.2},17.4,0.4,9.1,\textbf{19.4},HEIC (17.4),PKG (14.6)
7Z,Archive,0.9,1.4,\textbf{2.8},0.0,\textbf{53.2},0.0,4.6,0.1,HEIC (47.8),MOV (15.4)
BZ2,Archive,8.6,1.2,2.5,\textbf{13.9},75.8,4.3,74.8,\textbf{80.6},DMG (8.5),HEIC (3.2)
DEB,Archive,9.9,7.0,2.9,\textbf{13.8},1.5,0.0,\textbf{3.9},0.8,HEIC (38.8),RAR (10.6)
GZ,Archive,11.3,0.0,3.4,\textbf{13.2},\textbf{53.0},4.9,19.1,42.1,PKG (17.9),RPM (11.6)
PKG,Archive,\textbf{5.6},1.9,4.5,4.5,41.0,0.8,14.8,\textbf{62.7},GZ (6.5),APK (5.7)
RAR,Archive,13.0,4.2,3.0,\textbf{24.0},18.0,1.7,3.9,\textbf{46.6},ZIP (6.7),HEIC (6.3)
RPM,Archive,7.9,2.6,5.4,\textbf{13.6},10.7,0.5,8.9,\textbf{21.9},PKG (12.8),GZ (10.1)
XZ,Archive,0.0,\textbf{19.4},2.9,0.0,0.0,1.3,5.0,\textbf{12.1},HEIC (45.6),MOV (15.4)
ZIP,Archive,\textbf{16.2},1.8,9.5,13.8,\textbf{36.6},9.7,22.2,34.2,RAR (21.6),M4A (5.5)
EXE,Executable,\textbf{9.2},0.8,2.4,0.3,4.3,0.3,\textbf{5.4},2.7,HEIC (39.7),MOV (13.0)
MACH-O,Executable,\textbf{93.6},75.0,90.0,92.6,\textbf{95.7},71.8,93.7,95.0,DLL (2.1),TXT (0.6)
ELF,Executable,\textbf{87.6},65.4,83.7,85.9,\textbf{92.2},54.9,89.3,86.4,DLL (4.6),MACH-O(2.1)
DLL,Executable,\textbf{91.7},67.8,86.2,91.1,\textbf{94.4},65.7,86.4,91.7,TXT (2.8),ELF (0.6)
DOC,Office,\textbf{91.4},44.8,70.5,87.2,\textbf{91.6},12.4,77.6,88.9,PPT (2.2),APK (1.2)
DOCX,Office,14.7,0.7,6.3,\textbf{19.4},49.9,0.1,31.7,\textbf{60.6},EPUB (20.7),RAR (8.7)
KEY,Office,35.8,4.5,23.6,\textbf{44.1},\textbf{56.7},10.3,44.3,43.1,PPTX (19.9),PPT (13.2)
PPT,Office,\textbf{44.2},10.9,23.6,41.3,\textbf{58.4},15.7,32.5,53.5,PPTX (6.2),PKG (3.8)
PPTX,Office,38.3,14.4,24.0,\textbf{44.4},32.7,2.6,20.2,\textbf{36.0},PPT (14.6),KEY (11.5)
XLS,Office,\textbf{99.4},98.7,99.1,99.3,99.2,96.7,99.0,\textbf{99.3},DLL (0.3),MSI (0.1)
XLSX,Office,\textbf{96.8},65.6,92.7,95.2,\textbf{97.4},20.7,91.3,97.2,PKG (0.6),AI (0.5)
DJVU,Published,18.0,3.5,10.5,\textbf{24.6},34.8,\textbf{84.2},24.2,30.4,HEIC (33.7),MOV (12.5)
EPUB,Published,\textbf{29.3},2.0,9.2,20.7,74.3,3.3,56.0,\textbf{79.7},DOCX (10.5),PKG (2.0)
MOBI,Published,72.5,\textbf{72.8},71.5,72.4,\textbf{74.4},69.6,72.9,73.7,EPUB (19.2),DOCX (2.4)
PDF,Published,21.9,11.5,20.0,\textbf{23.1},\textbf{48.3},11.6,38.1,45.8,HEIC (13.1),MOV (4.7)
MD,Human-readable,\textbf{98.0},77.9,93.9,97.1,\textbf{99.9},86.6,98.9,97.4,HTML (1.7),TEX (0.5)
RTF,Human-readable,99.5,90.2,99.7,\textbf{99.7},100.0,87.7,\textbf{100.0},99.8,TXT ($<$0.1),DOC ($<$0.1)
TXT,Human-readable,95.0,85.9,\textbf{95.4},93.7,92.9,84.1,92.5,\textbf{93.0},DOCX (2.1),EPUB (2.0)
TEX,Human-readable,\textbf{97.8},66.3,93.0,97.6,\textbf{98.8},58.1,97.2,98.3,TXT (1.6),RTF ($<$0.1)
JSON,Human-readable,\textbf{99.8},84.3,99.5,99.5,\textbf{100.0},81.5,99.8,99.8,TXT ($<$0.1),HTML ($<$0.1)
HTML,Human-readable,\textbf{98.4},74.2,94.9,97.5,99.5,59.1,98.9,\textbf{99.6},TXT (0.2),MD (0.1)
XML,Human-readable,100.0,99.3,99.9,\textbf{100.0},100.0,99.9,100.0,\textbf{100.0},-,-
LOG,Human-readable,\textbf{100.0},99.3,99.9,99.9,99.9,91.2,99.9,\textbf{99.9},TXT ($<$0.1),TXT ($<$0.1)
CSV,Human-readable,\textbf{99.9},92.1,99.6,99.6,99.0,87.3,98.4,\textbf{99.7},TXT ($<$0.1),TXT ($<$0.1)
AIFF,Audio,95.7,96.2,94.9,\textbf{99.4},94.5,95.0,98.4,\textbf{98.8},TXT (0.8),TXT (0.8)
FLAC,Audio,66.9,18.5,13.2,\textbf{74.2},87.8,76.2,85.3,\textbf{97.9},ORF (0.4),ORF (0.4)
M4A,Audio,92.6,34.6,86.7,\textbf{94.3},96.5,29.6,\textbf{98.4},98.2,HEIC (0.3),MOV (0.3)
MP3,Audio,94.0,70.8,85.1,\textbf{94.4},98.9,77.6,98.8,\textbf{98.9},APK (0.2),APK (0.2)
OGG,Audio,84.2,55.3,72.3,\textbf{90.4},95.4,37.3,93.3,\textbf{95.5},FLAC (2.3),FLAC (2.3)
WAV,Audio,99.9,99.9,99.8,\textbf{100.0},74.0,58.2,97.6,\textbf{98.7},TXT (0.6),APK (0.4)
WMA,Audio,\textbf{99.2},73.8,98.7,98.2,99.9,95.7,99.9,\textbf{99.9},3GP ($<$0.1),APK ($<$0.1)
PCAP,Misc,\textbf{55.4},32.9,48.6,51.0,\textbf{95.6},69.6,94.7,95.4,HEIC (0.9),MOV (0.4)
TTF,Misc,98.1,86.8,96.3,\textbf{98.1},99.2,90.2,98.4,\textbf{99.3},DLL (0.3),DLL (0.3)
DWG,Misc,\textbf{96.5},90.1,93.0,96.2,\textbf{98.8},67.5,97.0,96.9,BMP (0.4),BMP (0.4)
SQLITE,Misc,\textbf{98.9},87.2,97.1,98.6,99.5,91.3,98.6,\textbf{99.8},DLL ($<$0.1),DLL ($<$0.1)
-,\textbf{Total Wins},\textbf{32},\textbf{2},\textbf{3},\emph{\textbf{38}},\textbf{27},\textbf{1},\textbf{4},\emph{\textbf{43}},-,-
\end{filecontents*}
\pgfplotstabletypeset[
    column type=c,
    every head row/.style={
        before row={
            \toprule
            \multicolumn{2}{c}{\textbf{}} & \multicolumn{4}{c}{\textbf{512 bytes}} & \multicolumn{4}{c}{\textbf{4096 bytes}} & \multicolumn{2}{c}{\textbf{Misclassification}} \\
        },
        after row=\midrule,
    },
    col sep = comma,
    string type,
    string replace*={_}{\textsubscript},
    every last row/.style={after row=\bottomrule  \multicolumn{6}{l}{Note: Ties were broken depending upon runtime speed.}},
    assign column name/.style={/pgfplots/table/column name={\textbf{#1}}},
    display columns/0/.style={column type={p{.065\textwidth}}},
    display columns/1/.style={column type={p{.11\textwidth}}},
    columns/Co_occur_512/.style = {column name=NN-CO, column type={r}},
    columns/Feature_512/.style = {column name=NN-GF, column type={r}},
    columns/Sceadan_512/.style = {column name=Sceadan, column type={r}},
    columns/CNN_512/.style = {column name=FiFTy, column type={r}},
    columns/Co_occur_4k/.style = {column name=NN-CO, column type={r}},
    columns/Feature_4k/.style = {column name=NN-GF, column type={r}},
    columns/Sceadan_4k/.style = {column name=Sceadan, column type={r}},
    columns/CNN_4k/.style = {column name=FiFTy, column type={r}},
    columns/top1/.style = {column name=Top 1, column type={p{.085\textwidth}}},
    columns/top2/.style = {column name=Top 2, column type={p{.095\textwidth}}}
    ]
    {data/misclassification.csv}%
    }
\end{table*}
\bibliographystyle{IEEEbib-abbrev}
\bibliography{./bib/references}

\begin{IEEEbiography}[{\includegraphics[width=1in,height=1.25in,clip,keepaspectratio]{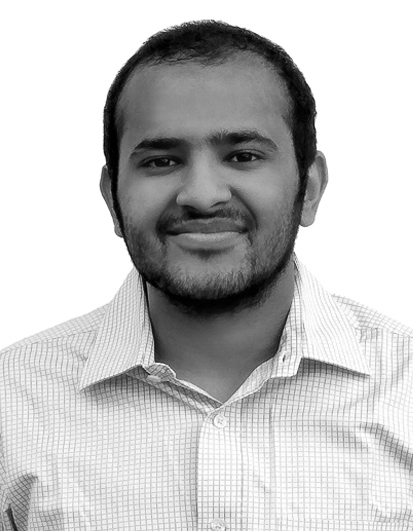}}]{Govind Mittal} earned his Bachelor of Engineering in Computer Science and Master of Science in Mathematics from Birla Institute of Technology and Science (BITS) Pilani, India. He is currently pursuing the Ph.D. degree with the Department of Computer Science and Engineering, New York University Tandon School of Engineering. His research interests includes digital forensics, tackling disinformation and machine learning for cyber security. 
\end{IEEEbiography}

\begin{IEEEbiography}[{\includegraphics[width=1in,height=1.25in,clip,keepaspectratio]{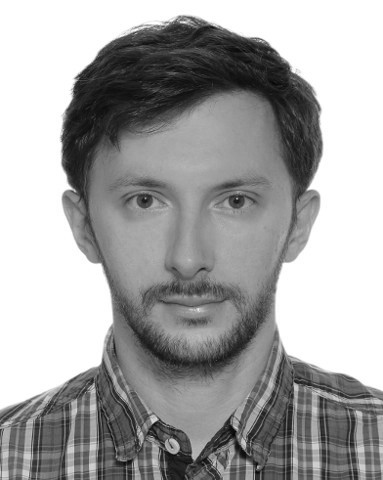}}]{Paweł Korus} received his M.Sc. and Ph.D. degrees in telecommunications (with honors) from the AGH University of Science and Technology in 2008, and in 2013, respectively. Since 2014 he has been an assistant professor with the Department of Telecommunications, AGH University of Science and Technology, Krakow, Poland. He did his postdoctoral research at the College of Information Engineering, Shenzhen University, China. He is currently a visiting professor at the Tandon School of Engineering, New York University, USA.

His research interests include various aspects of multimedia security, image processing, and low-level vision, with particular focus on content authentication and protection techniques for digital photographs. In 2015 he received a scholarship for outstanding young scientists from the Polish Ministry of Science and Higher Education. 
\end{IEEEbiography}

\begin{IEEEbiography}[{\includegraphics[width=1in,height=1.25in,clip,keepaspectratio]{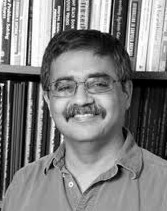}}]{Nasir Memon} is a professor in the Department of Computer Science and Engineering at NYU Tandon School of Engineering. He is one of the founding members of the Center for Cyber Security (CCS) , a collaborative initiative of multiple schools within NYU including.  He is the founder of CSAW, and  the Offensive Security, Incident Response and Internet Security Laboratory (OSIRIS) lab at NYU Tandon. His research interests include digital forensics, biometrics, data compression, network security and human behavior. Memon earned a Bachelor of Engineering in Chemical Engineering and a Master of Science in Mathematics from Birla Institute of Technology and Science (BITS) in Pilani, India. He received a PhD in Computer Science from the University of Nebraska.

Professor Memon has published over 250 articles in journals and conference proceedings and holds a dozen patents in image compression and security. He has won several awards including the Jacobs Excellence in Education award and several best paper awards. He has been on the editorial boards of several journals and was the Editor-In-Chief of Transactions on Information Security and Forensics. He is an IEEE Fellow and an SPIE fellow. 
\end{IEEEbiography}

\end{document}


\begin{center}
    {\Large\textbf{Supplementary Materials for \\ \vspace{8pt} FiFTy: Large-scale File Fragment Type Identification using Neural Networks}}

    \vspace{8pt} Govind Mittal$^1$, Paweł Korus$^{1,2}$, and Nasir Memon$^{1}$ 
    
    New York University$^1$, AGH University of Science and Technology$^2$
\end{center}

\markboth{IEEE TRANSACTIONS ON INFORMATION FORENSICS AND SECURITY}{}

\vspace{20pt}

\section{Source Code}

Our file fragment classifier toolbox is available at \url{https://github.com/mittalgovind/fifty} and the supporting FFT-75 dataset is available at \url{http://dx.doi.org/10.21227/kfxw-8084}

\section{Contents}

\begin{itemize}
    \itemsep2pt
    \item Effect of hyper-parameter variation on various scenarios with block size of 512 and 4096 bytes.
    \item Full confusion matrices for block sizes of 512 and 4096 bytes of :
      \begin{itemize}
        \item FiFTy 
        \item Sceadan
        \item NN-GF
        \item NN-CO
        \item Scenario \#2 - \#6
      \end{itemize}
    \item List of sources for each filetype in the corpus FFT-75.
\end{itemize}

\newpage
\begin{landscape}
 \begin{figure}[]
  \centering
  \caption{Impact of hyper-parameter variation on the validation accuracy for all scenarios with 512-byte inputs: (top) violin plots showing distribution of accuracy for sampled networks - dark color corrosponds to the TPE chosen value; (bottom) \emph{expected improvement} estimates from TPE.}
  \includegraphics[width=1.25\textwidth]{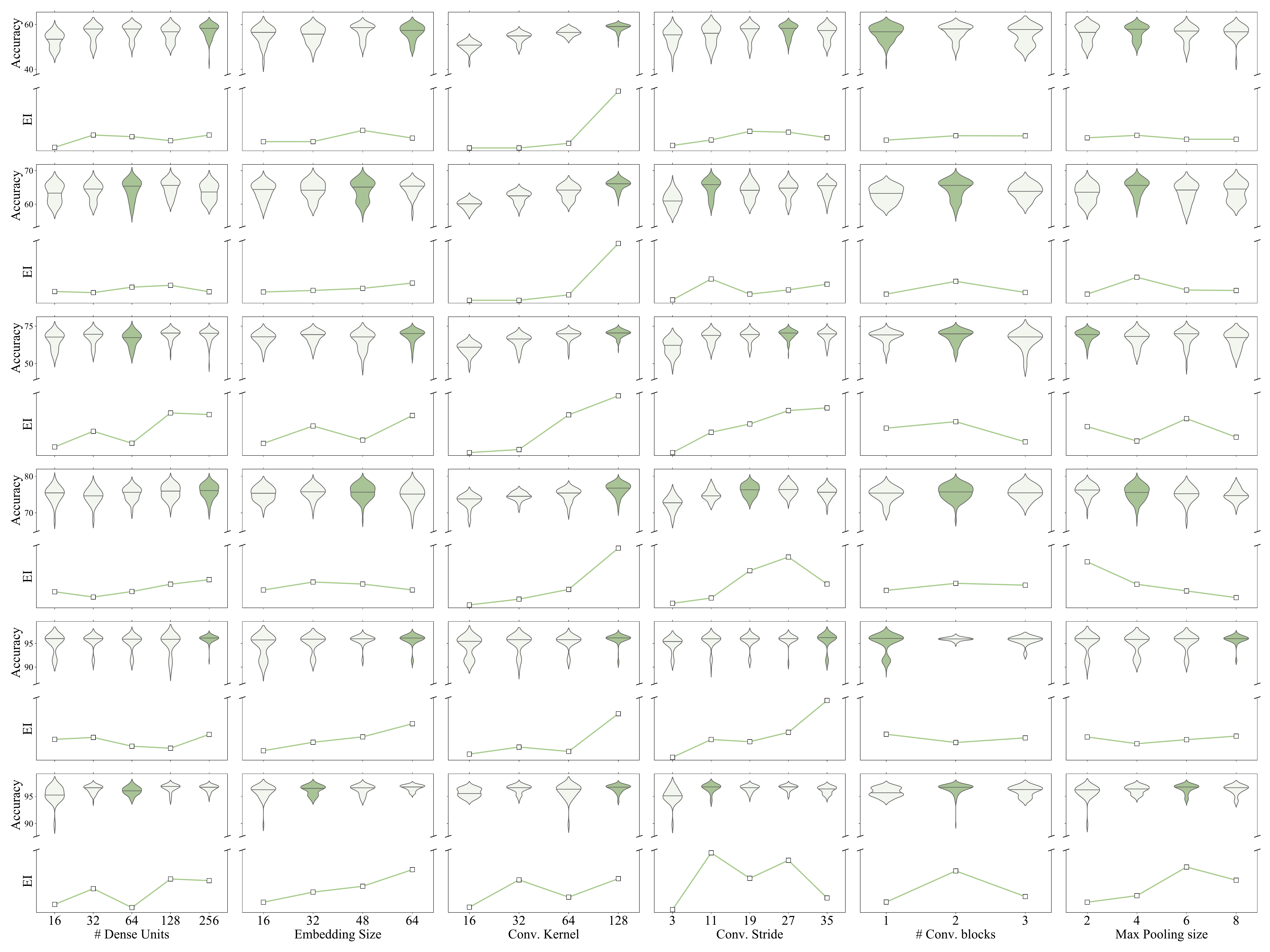}
  \label{fig:hyper_512}
 \end{figure}
\end{landscape}

\begin{landscape}
  \begin{figure}[]
   \centering
   \caption{Impact of hyper-parameter variation on the validation accuracy for all scenarios with 4096-byte inputs: (top) violin plots showing distribution of accuracy for sampled networks - dark color corrosponds to the TPE chosen value; (bottom) \emph{expected improvement} estimates from TPE.}
   \includegraphics[width=1.25\textwidth]{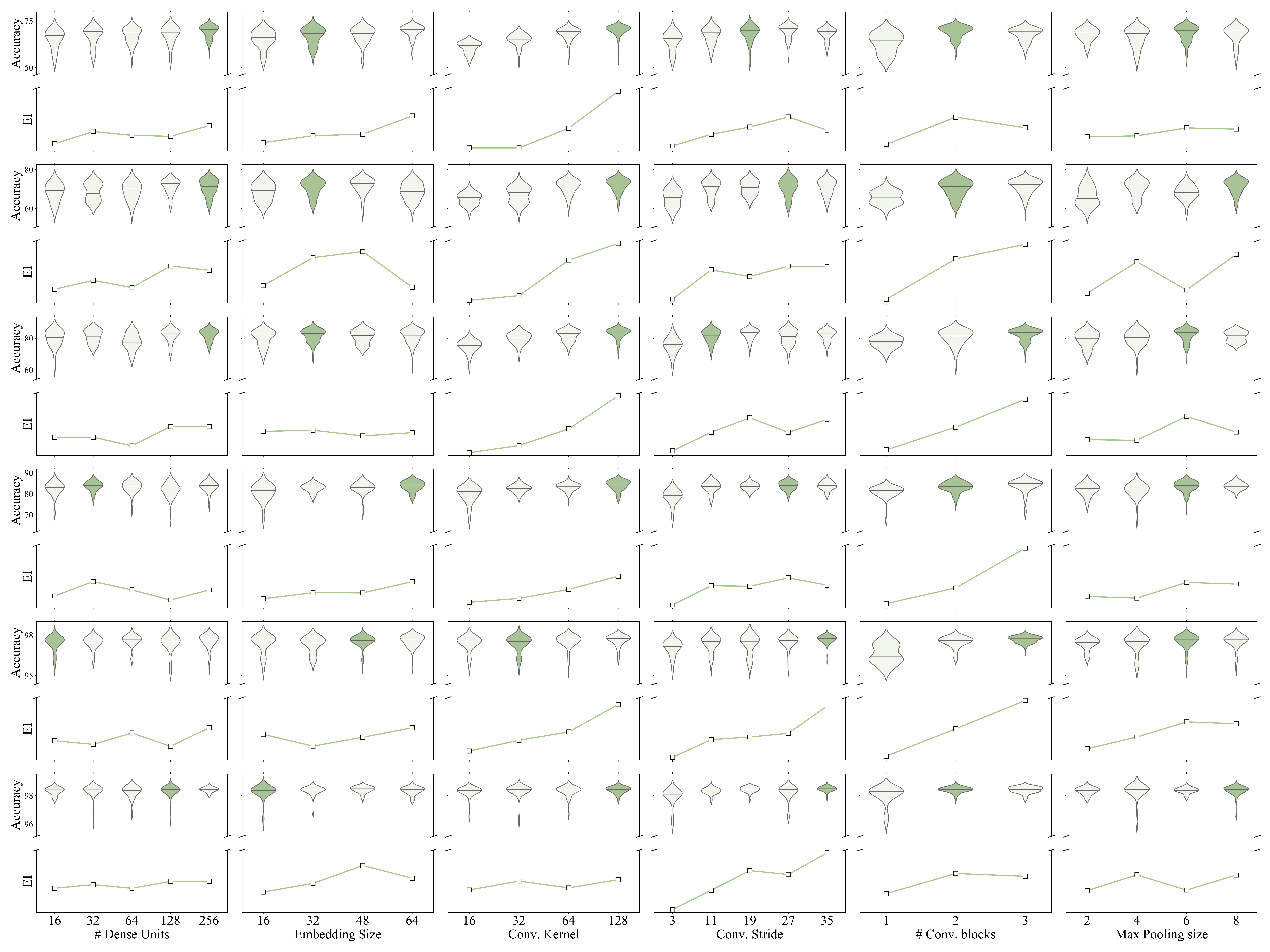}
   \label{fig:hyper_4k}
  \end{figure}
 \end{landscape}

\begin{figure*}
  \includegraphics[width=1.0\textwidth]{cnf_mats/full/512_1_cnf_mat.pdf}
  \caption{Confusion matrix for \emph{FiFTy} with block size of 512 bytes. Darker color means higher value and dot $(.)$ means a non-zero value greater than 0.2\%.}
\end{figure*}

\begin{figure*}
  \includegraphics[width=1.0\textwidth]{cnf_mats/full/4k_1_cnf_mat.pdf}
  \caption{Confusion matrix for \emph{FiFTy} with block size of 4096 bytes. Darker color means higher value and dot $(.)$ means a non-zero value greater than 0.2\%.}
\end{figure*}

\begin{figure*}
  \includegraphics[width=1.0\textwidth]{cnf_mats/full/sceadan_512_1_cnf_mat.pdf}
  \caption{Confusion matrix for Sceadan with block size of 512 bytes. Darker color means higher value and dot $(.)$ means a non-zero value greater than 0.2\%.}
\end{figure*}

\begin{figure*}
  \includegraphics[width=1.0\textwidth]{cnf_mats/full/sceadan_4k_1_cnf_mat.pdf}
  \caption{Confusion matrix for Sceadan with block size of 4096 bytes. Darker color means higher value and dot $(.)$ means a non-zero value greater than 0.2\%.}
\end{figure*}

\begin{figure*}
  \includegraphics[width=1.0\textwidth]{cnf_mats/full/feat_512_1_cnf_mat.pdf}
  \caption{Confusion matrix for baseline NN-GF with block size of 512 bytes. Darker color means higher value and dot $(.)$ means a non-zero value greater than 0.2\%.}
\end{figure*}

\begin{figure*}
  \includegraphics[width=1.0\textwidth]{cnf_mats/full/feat_4k_1_cnf_mat.pdf}
  \caption{Confusion matrix for baseline NN-GF with block size of 4096 bytes. Darker color means higher value and dot $(.)$ means a non-zero value greater than 0.2\%.}
\end{figure*}

\begin{figure*}
  \includegraphics[width=1.0\textwidth]{cnf_mats/full/co_occur_512_1_cnf_mat.pdf}
  \caption{Confusion matrix for baseline NN-CO with block size of 512 bytes. Darker color means higher value and dot $(.)$ means a non-zero value greater than 0.2\%.}
\end{figure*}

\begin{figure*}
  \includegraphics[width=1.0\textwidth]{cnf_mats/full/co_occur_4k_1_cnf_mat.pdf}
  \caption{Confusion matrix for baseline NN-CO with block size of 4096 bytes. Darker color means higher value and dot $(.)$ means a non-zero value greater than 0.2\%.}
\end{figure*}

\begin{figure*}
  \subfloat[][Scenario \#2]{\includegraphics[width=0.5\textwidth]{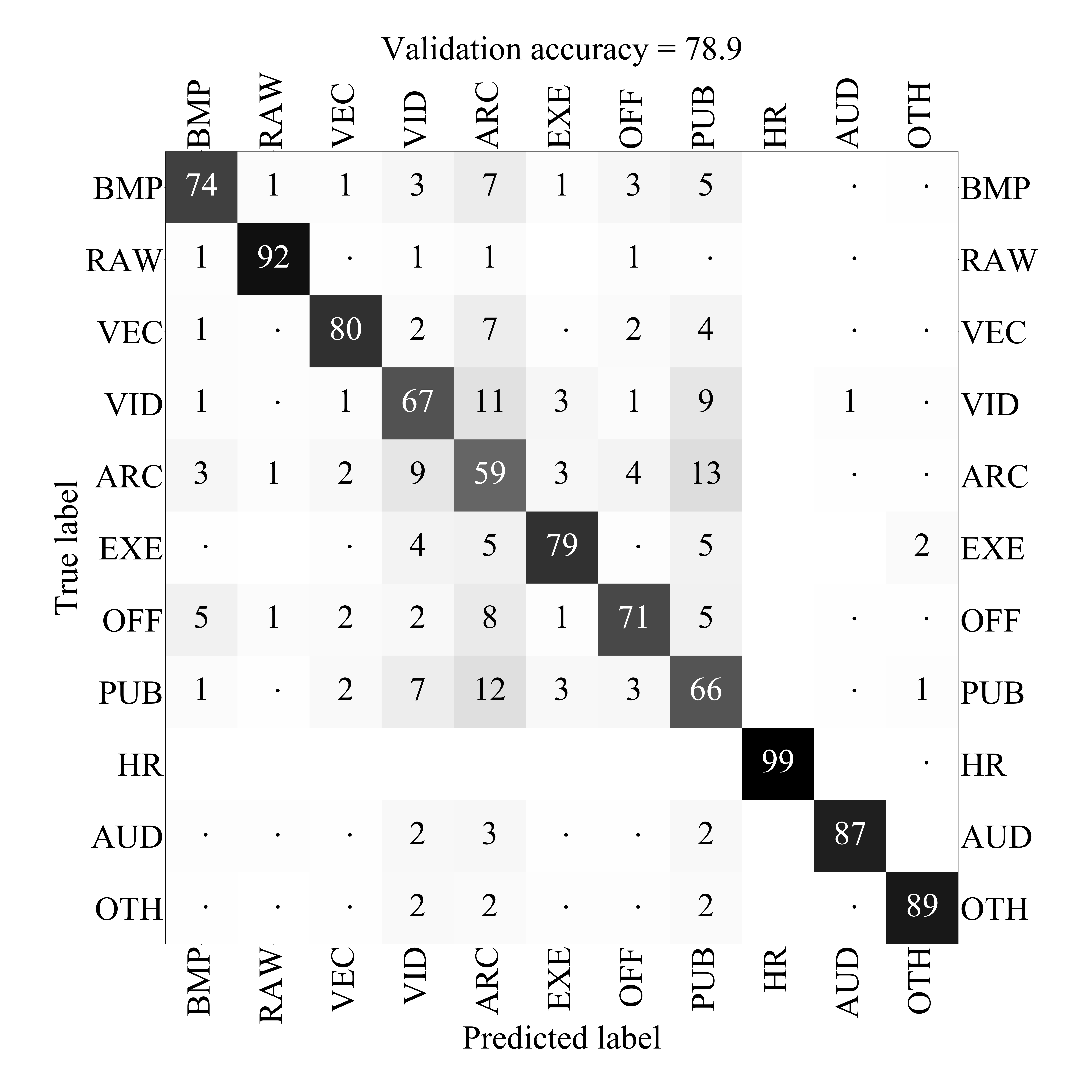}} 
  \subfloat[][Scenario \#3]{\includegraphics[width=0.5\textwidth]{cnf_mats/full/512_3_cnf_mat.pdf}} \\
  \subfloat[][Scenario \#4]{\includegraphics[width=0.5\textwidth]{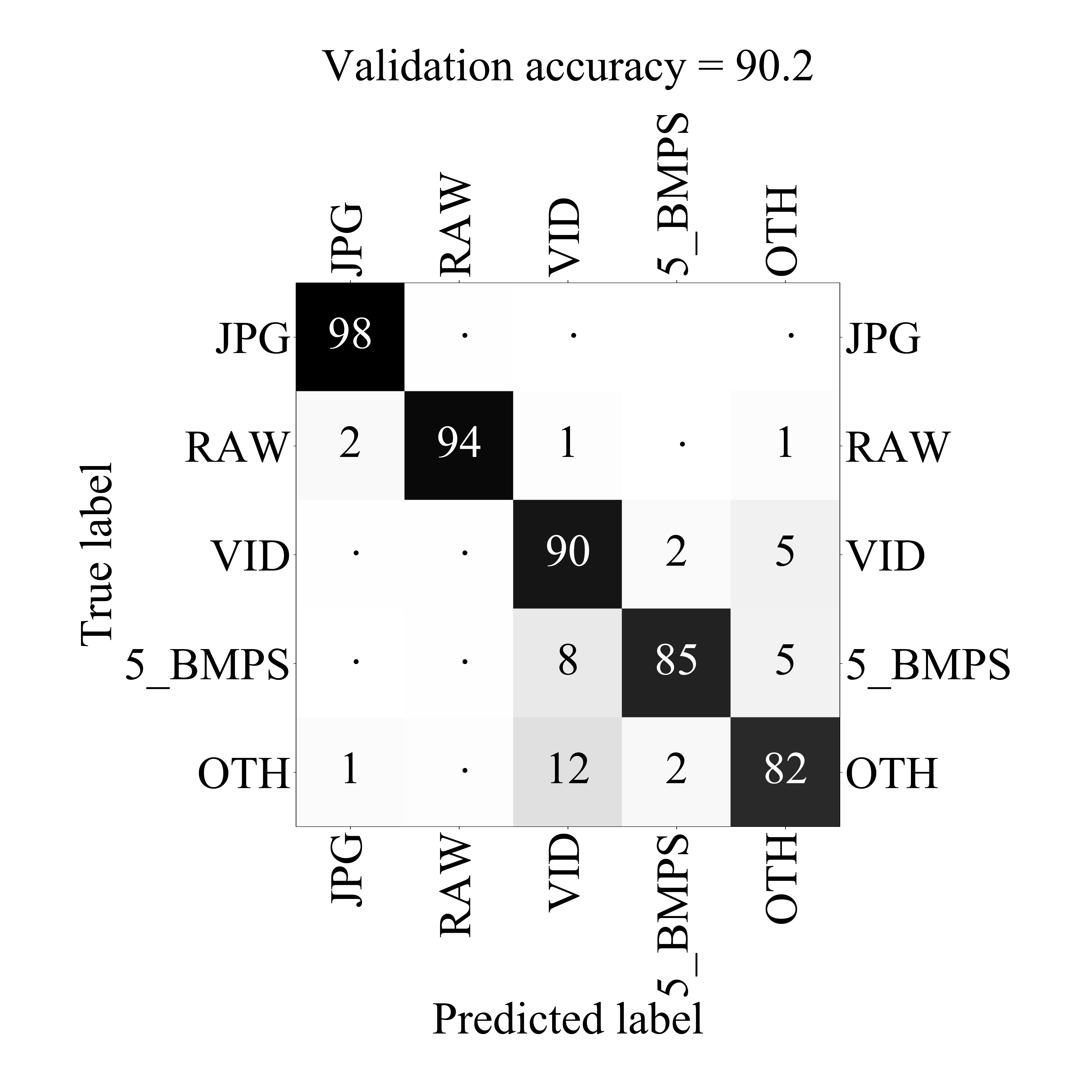}} 
    \subfloat[][Scenario \#5]{\includegraphics[width=0.25\textwidth]{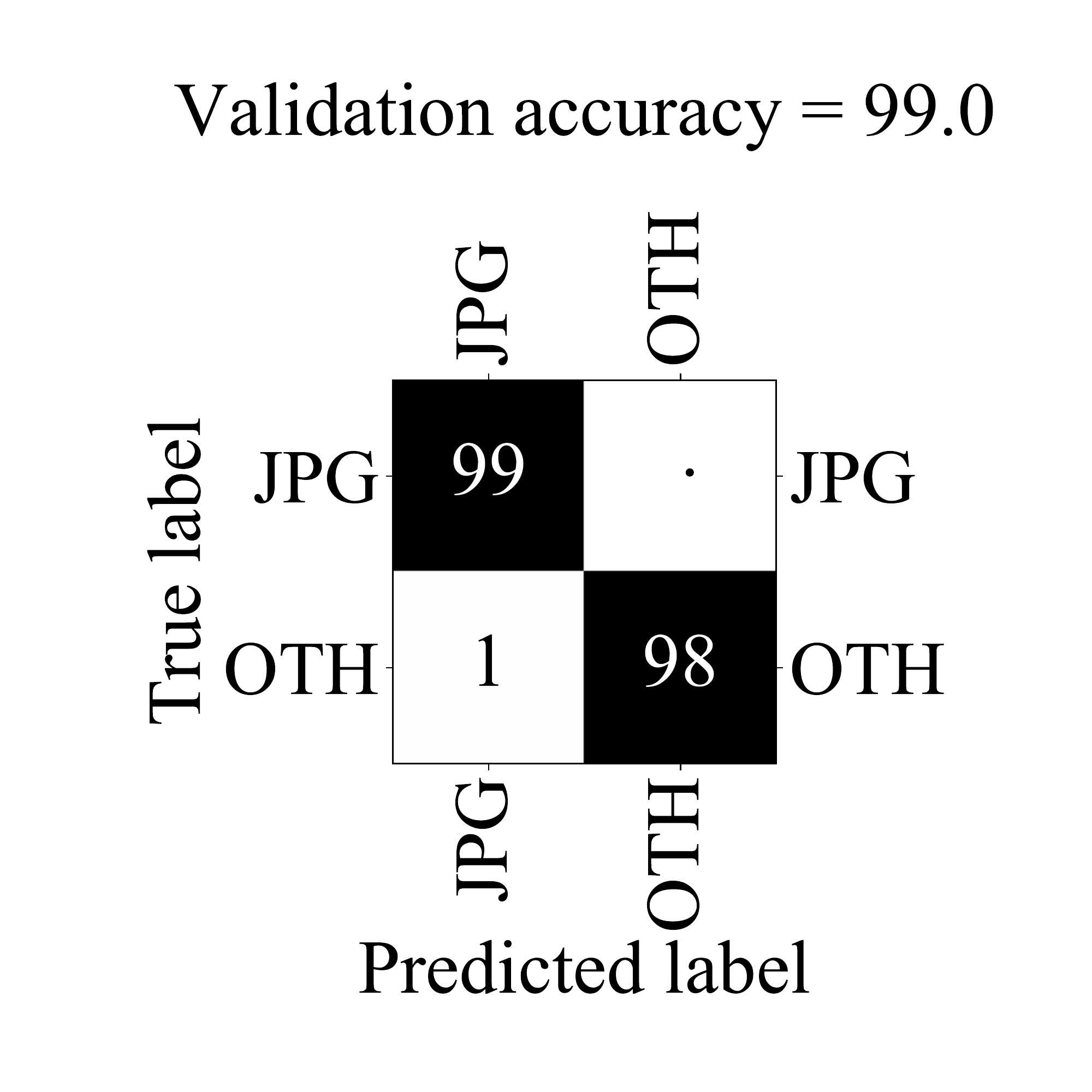}}
    \subfloat[][Scenario \#6]{\includegraphics[width=0.25\textwidth]{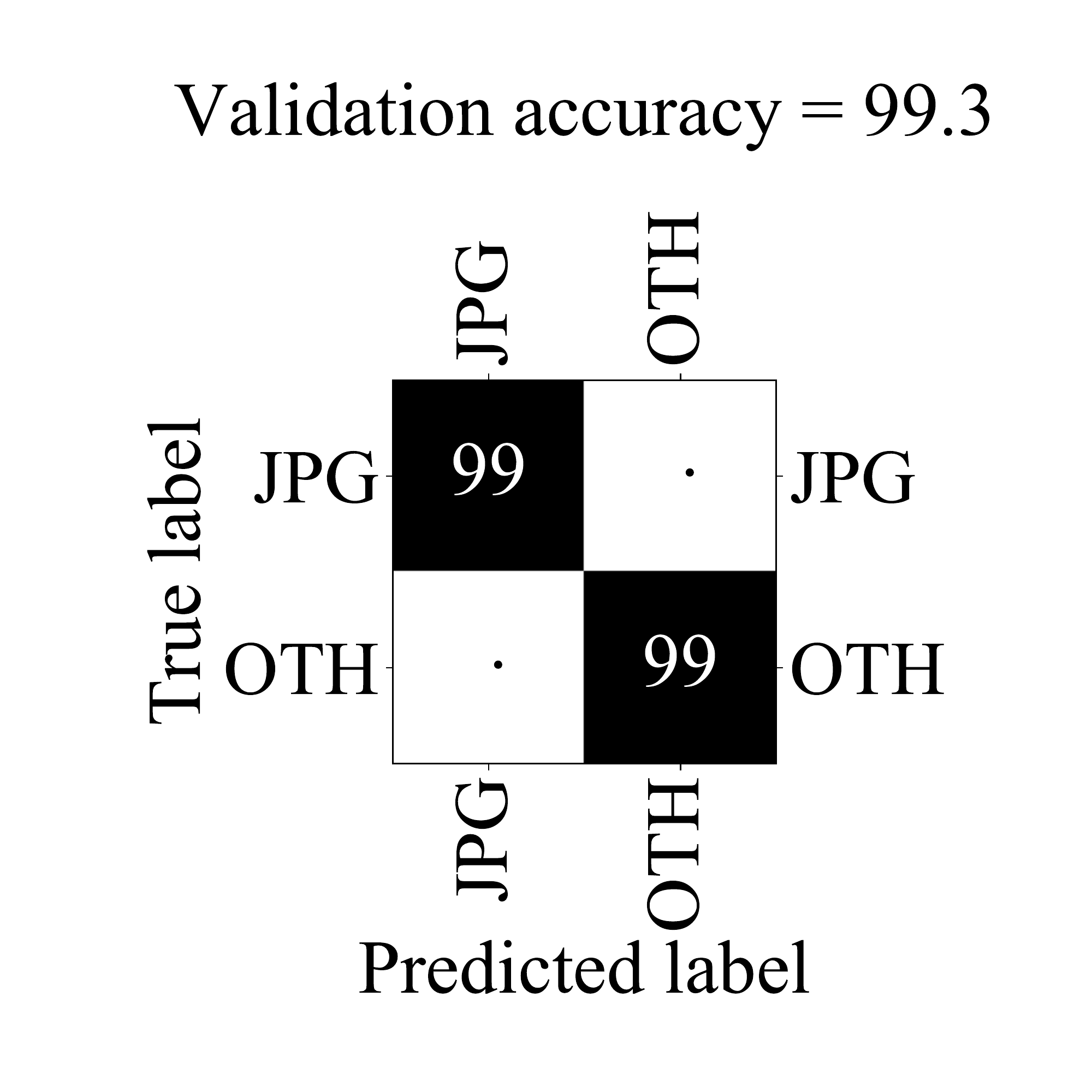}}
  \caption{Confusion matrices of the remaining five scenarios with block size of 512 bytes. Darker color means higher value and dot $(.)$ means a non-zero value greater than 0.2\%.}
  \label{}
\end{figure*}

\begin{figure*}
  \subfloat[][Scenario \#2]{\includegraphics[width=0.5\textwidth]{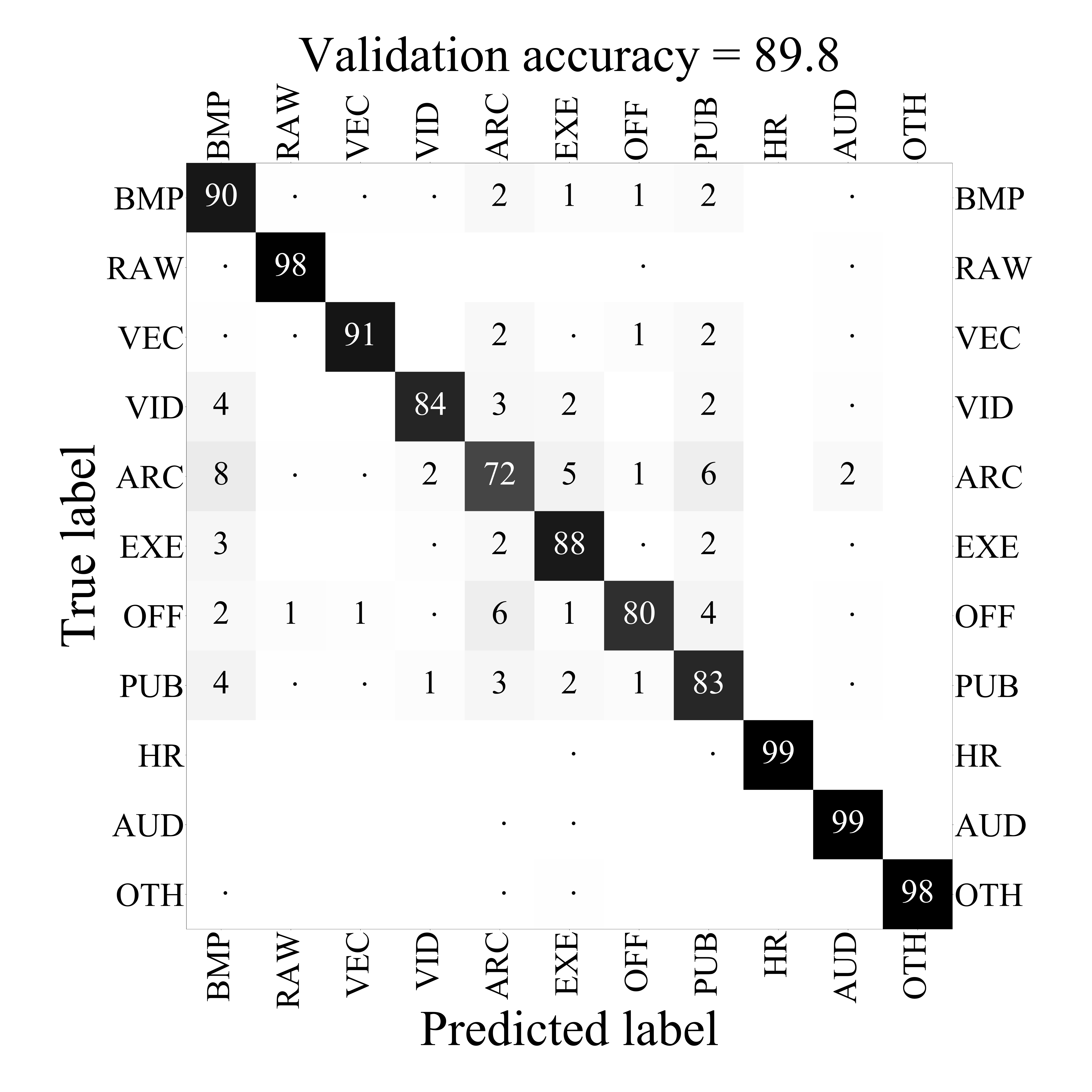}} 
  \subfloat[][Scenario \#3]{\includegraphics[width=0.5\textwidth]{cnf_mats/full/4k_3_cnf_mat.pdf}} \\
  \subfloat[][Scenario \#4]{\includegraphics[width=0.5\textwidth]{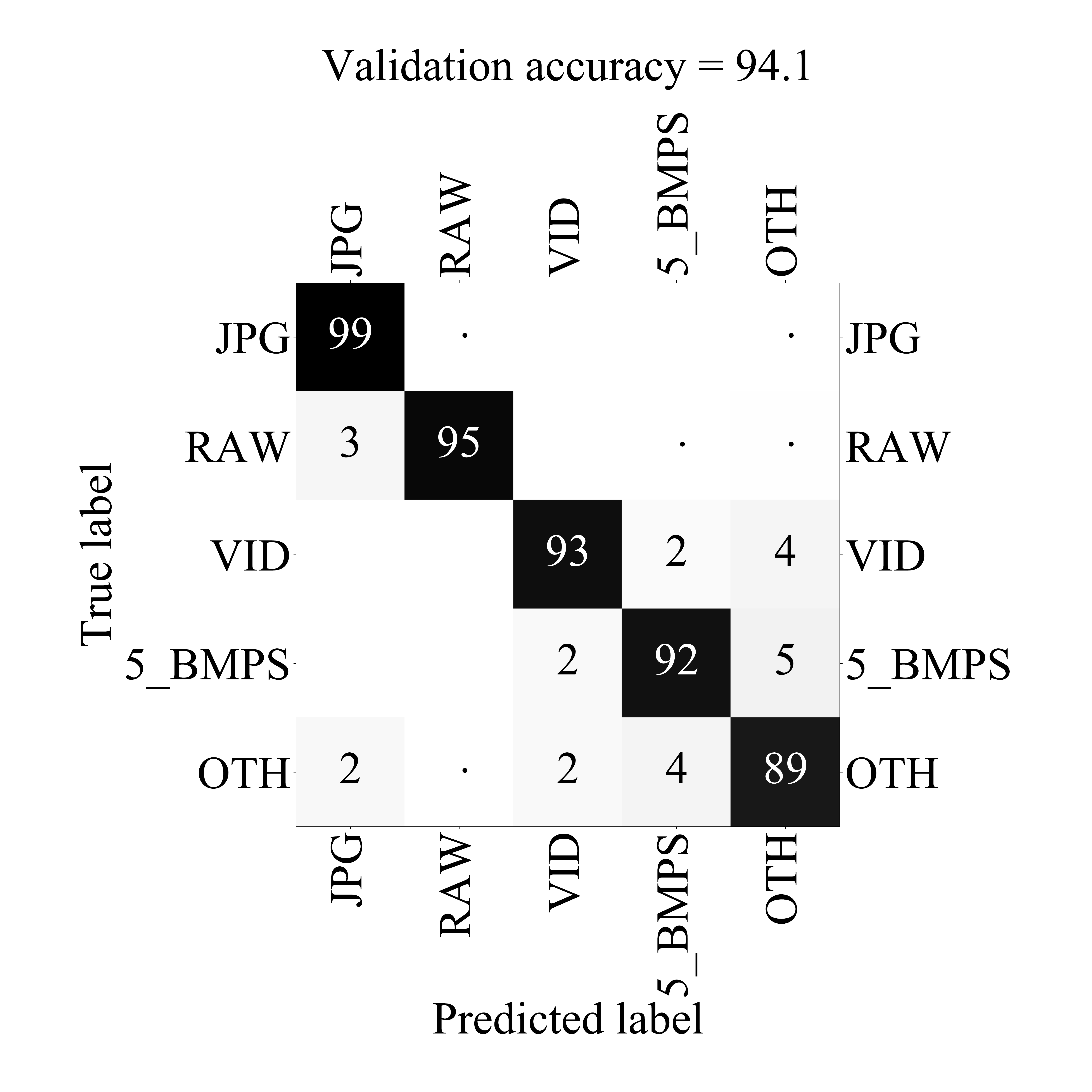}} 
    \subfloat[][Scenario \#5]{\includegraphics[width=0.25\textwidth]{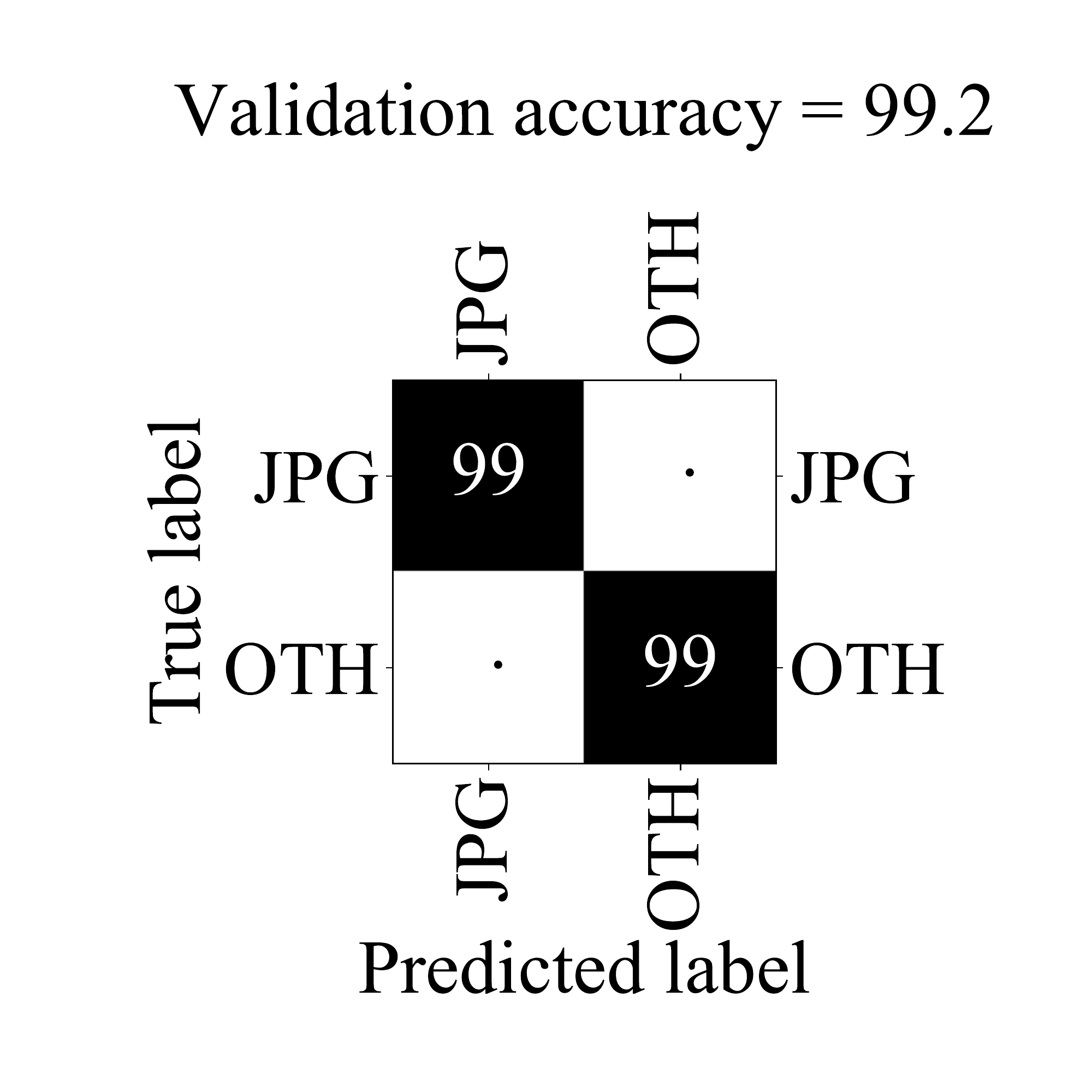}}
    \subfloat[][Scenario \#6]{\includegraphics[width=0.25\textwidth]{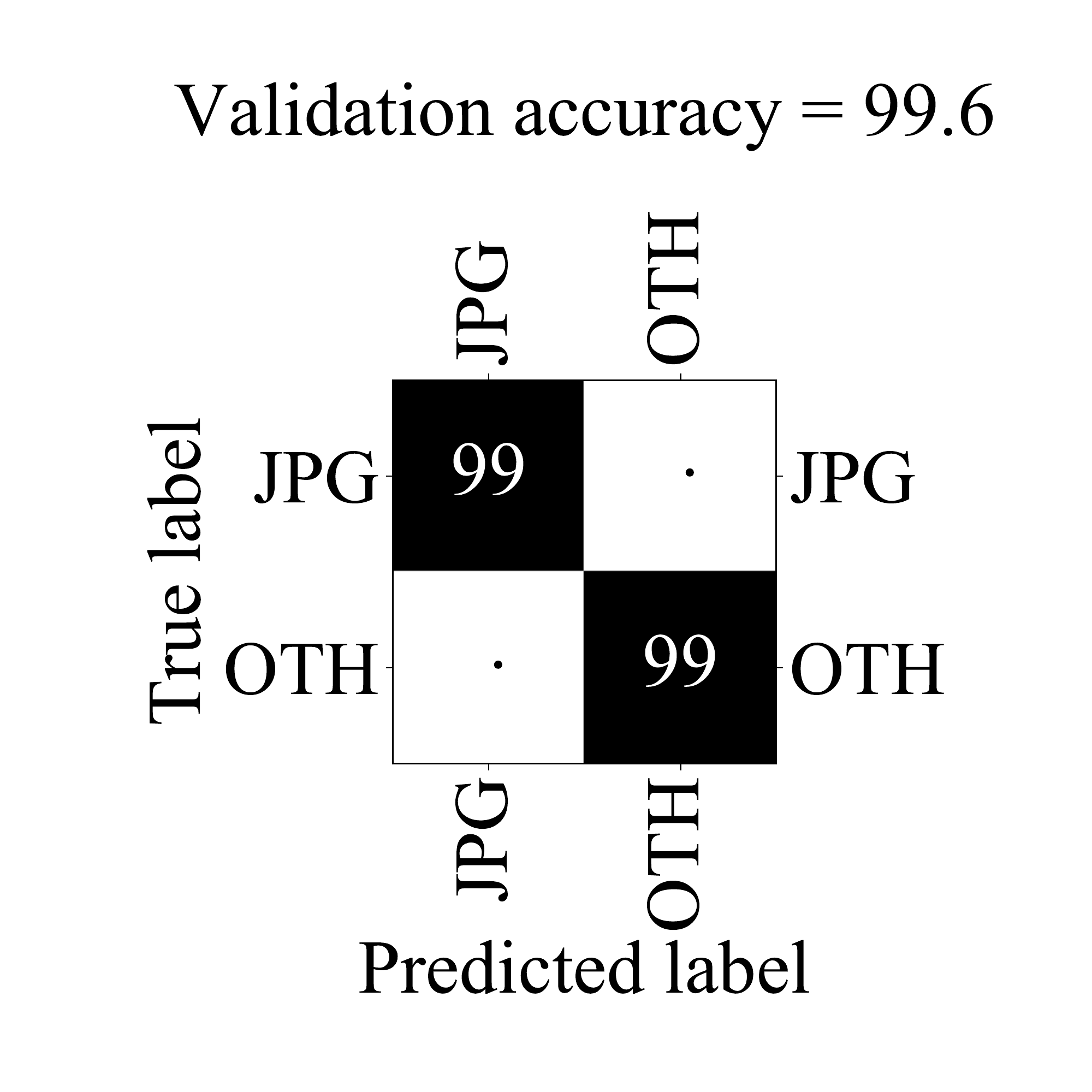}}
  \caption{Confusion matrices of the remaining five scenarios with block size of 4096 bytes. Darker color means higher value dot $(.)$ means a non-zero value greater than 0.2\%.}
  \label{}
\end{figure*}

\begin{table*}[p]
    \caption{File Types Studied along with their description and sources$^1$.}
    \resizebox{0.44\textwidth}{!}{%
\parbox[t]{0.45\textwidth}{
\begin{tabular}{C{1cm}L{3.7cm}L{3.3cm}}
    	\toprule
    	 \textbf{Filetype} & \textbf{Description} & \textbf{Source(s)} \\ \midrule
	 	 JPG & Joint Photographers Experts Group (JPEG) & dpreview.com, self-shot \\ \midrule
		 ARW & Raw Sony camera images & wesaturate.com \\ \midrule
		 CR2 & Raw Canon camera images & wesaturate.com \\ \midrule
		 DNG & Raw Adobe camera images & wesaturate.com \\ \midrule
		 GPR & Raw GoPro camera images & self-shot with GoPro Hero 6 Black \\ \midrule
		 NEF & Raw Nikon camera images & loki.disi.unitn.it/raise \\ \midrule
		 NRW & Raw Nikon camera images & photographyblog.com \\ \midrule
		 ORF & Raw Olympus camera images & rawsamples.ch/index.php \\ \midrule
		 PEF & Raw Pentax camera images & dpreview.com \\ \midrule
		 RAF & Raw Fuji camera images & rawsamples.ch/index.php \\ \midrule
		 RW2 & Raw Panasonic camera images & rawsamples.ch/index.php \\ \midrule
		 3FR & Raw Hasselblad camera images & hasselblad.com \\ \midrule
		 TIFF & Tagged Image File Format & rawsamples.ch/index.php \\ \midrule
		 HEIC & High Efficiency Image Format based on video frames & Encoded from JPG images from Nokiatech \\ \midrule
		 BMP & Bitmap images & 3axis.co/free-bmp-files \\ \midrule
		 GIF & Graphic Interchange Format & commons.wikimedia.org \\ \midrule
		 PNG & Portable Network Graphics & commons.wikimedia.org, archive.org \\ \midrule
		 AI & Adobe Illustrator vector image & 3axis.co/free-ai-files \\ \midrule
		 EPS & Encapsulated PostScript vector & free-vectors.com \\ \midrule
		 PSD & Photoshop vector file & livven.me/psds/ \\ \midrule
		 MOV & QuickTime File Format & self-shot98 \\ \midrule
		 MP4 & MPEG-4 Part 14 & archive.org, self-shot \\ \midrule
		 3GP & Multimedia container videos format & archive.org \\ \midrule
		 AVI & Audio Video Interleave container format & archive.org \\ \midrule
		 MKV & Matroska Multimedia Container & archive.org \\ \midrule
		 OGV & Ogg Vorbis video encoding format & commons.wikimedia.org \\ \midrule
		 WEBM & Web videos & commons.wikimedia.org \\ \midrule
		 APK & Android application package & archive.org \\ \midrule
		 EXE & Windows executable & installers taken from open source developers' website \\ \midrule
		 JAR & Java class package (compiled) & jar-download.com \\ \midrule
		 MSI & Windows Installer & installers taken from open source developers' website \\ \midrule
		 DMG & macOS application package & installers taken from open source developers' website \\ \midrule
		 7Z & 7-zip archive & rawsamples.ch \\ \midrule
		 BZ2 & Burrows–Wheeler archive & installers taken from open source developers' website \\ \midrule
		 DEB & Linux/Unix application package & installers taken from open source developers' website \\ \midrule
		 GZ & GNU Gzip & installers taken from open source developers' website \\ \midrule
		 PKG & macOS compressed installer & installers taken from open source developers' website \\ \midrule
		 TTF & True-type font & dafont.com \\
    	 \bottomrule
    	 \multicolumn{3}{l}{$^1$All the links in the table have been accessed on or before 31st March, 2019.} \\
    \end{tabular}
    }%
    }
    \hspace{14mm}
    \resizebox{0.44\textwidth}{!}{%
    \parbox[t]{0.44\columnwidth}{
    \begin{tabular}{C{12mm}L{3.65cm}L{3.3cm}}
    	\toprule
    	 \textbf{Filetype} & \textbf{Description} & \textbf{Source(s)} \\ \midrule
		 RAR & Roshal Archive by Microsoft & archive.org \\ \midrule
    	 RPM & RPM package manager (Red Hat) & installers taken from open source developers' website \\ \midrule
		 XZ & XZ (GNU LGPL/GPL) & installers taken from open source developers' website \\ \midrule
		 ZIP & ZIP archive & self-compressed \\ \midrule
		 DJVU & Digital Document Format by Yann LeCun & commons.wikimedia.org \\ \midrule
		 EPUB & Electronic Publication for iBooks & gutenberg.org \\ \midrule
		 MOBI & Kindle E-book & gutenberg.org \\ \midrule
		 DOC & Microsoft Office (2007) Word  & converted from .epub files \\ \midrule
		 DOCX & Microsoft Office (2013) Word  & gutenberg.org \\ \midrule
		 MD & Markdown & converted from HTML, github repositories \\ \midrule
		 RTF & Rich text format & converted from .txt files \\ \midrule
		 TXT & Text file & gutenberg.org \\ \midrule
		 PDF & Portable Document Format & arXiv.org, gutenberg.org, commons.wikimedia.org \\ \midrule
		 KEY & macOS keynote presentation  & converted from .pptx files\\ \midrule
		 PPT & Microsoft Office (2007) Powerpoint  & kmworld.com \\ \midrule
		 PPTX & Microsoft Office (2013) Powerpoint  & pptx-templates.com, kmworld.com \\ \midrule
		 LOG & Log files & self syslogs \\ \midrule
		 JSON & JavaScript Object Notation for database & kaggle.com, github.com \\ \midrule
		 DWG & CAD drawing & dwgmodels.com \\ \midrule
		 SQLITE & SQL database & kaggle.com \\ \midrule
		 CSV & Comma-separated values & kaggle.com, who.int \\ \midrule
		 XLS & Microsoft Office (2007) Excel & converted from .csv files \\ \midrule
		 XLSX & Microsoft Office (2013) Excel & converted from .csv files \\ \midrule
		 AIFF & Audio Interchange File Format & Converted from .flac files \\ \midrule
		 FLAC & Free Lossless Audio Codec & commons.wikimedia.org, converted from .wav files \\ \midrule
		 M4A & Audio-only MPEG-4 & freemusicarchive.org \\ \midrule
		 MP3 & MPEG-1/2 Audio Layer III & freemusicarchive.org \\ \midrule
		 OGG & Audio container format developed by Xiph.Org & commons.wikimedia.org \\ \midrule
		 WAV & Waveform Audio File format & commons.wikimedia.org \\ \midrule
		 WMA & Windows Media Audio developed by Microsoft & converted from .wav files \\ \midrule
		 HTML & HyperText Markup Language & waybackmachine.org \\ \midrule
		 XML & Extensible Markup Language & gutenberg.org \\ \midrule
		 PCAP & Wireshark captured network packets & Capture-the-flag events from github.com \\ \midrule
		 DLL & Dynamic Link Library (Windows Executable) & dlldump.com \\ \midrule
		 ELF & Linux executable & Capture-the-flag events from github.com \\ \midrule
		 MACH-O & macOS executable & Anaconda virtual env. \\ \midrule
		 TEX & \LaTeX & arXiv.org \\ \bottomrule
 		\\
		 
    \end{tabular}
	}%
	}
    
    \label{tab:sources}
\end{table*}